# Two unseen massive white dwarf candidates in close binaries

**Yuta Shiraishi,[1,2,*] Kenta Hotokezaka,[2,3] Kento Masuda,[4] Satoshi Honda,[5] Ataru Tanikawa,[6] Soetkin Janssens,[2] Takato Tokuno,[7,8] Takumi Shimasue,[1,2] Ryoga Honjo,[1,2] Bun'ei Sato,[9] Masashi Omiya,[10,11] Akito Tajitsu,[12] and Hideyuki Izumiura[12]**

[1]Department of Physics, Graduate School of Science, The University of Tokyo, 7-3-1 Hongo, Bunkyo-ku, Tokyo 113-0033, Japan
[2]Research Center for the Early Universe, Graduate School of Science, The University of Tokyo, 7-3-1 Hongo, Bunkyo-ku, Tokyo 113-0033, Japan
[3]Max Planck Institute for Gravitational Physics (Albert Einstein Institute), Am Mühlenberg 1, Potsdam-Golm 14476, Germany
[4]Department of Earth and Space Science, Graduate School of Science, Osaka University, 1-1 Machikaneyama-cho, Toyonaka, Osaka 560-0043, Japan
[5]Nishi-Harima Astronomical Observatory, Center for Astronomy, University of Hyogo, 407-2 Nishigaichi, Sayo-cho, Sayo, Hyogo 679-5313, Japan
[6]Center for Information Science, Fukui Prefectural University, 4-1-1 Matsuoka Kenjojima, Eiheiji-cho, Fukui, Fukui 910-1195, Japan
[7]Department of Astronomy, Graduate School of Science, The University of Tokyo, 7-3-1 Hongo, Bunkyo-ku, Tokyo 113-0033, Japan
[8]School of Arts & Sciences, The University of Tokyo, 3-8-1 Komaba, Meguro-ku, Tokyo 153-8902, Japan
[9]Department of Earth and Planetary Sciences, School of Science, Institute of Science Tokyo, 2-12-1 Ookayama, Meguro-ku, Tokyo 152-8551, Japan
[10]Astrobiology Center, National Institutes of Natural Sciences, 2-21-1 Osawa, Mitaka, Tokyo 181-8588, Japan
[11]National Astronomical Observatory of Japan, 2-21-1 Osawa, Mitaka, Tokyo 181-8588, Japan
[12]Subaru Telescope Okayama Branch Office, National Astronomical Observatory of Japan, 3037-5 Honjou, Kamogata, Asakuchi, Okayama 719-0232, Japan

*shiraishi-yuta999@g.ecc.u-tokyo.ac.jp

## Abstract

We report the discovery of two binary systems, each consisting of a slightly bloated G-type main-sequence star and an unseen companion, identified through photometric data from TESS and radial velocity variation from Gaia. High-resolution spectroscopy confirms orbital periods of 1.37 and 2.67 d with circular orbits. The visible components have masses of $\sim$0.9–1.0 $M_\odot$, while the minimum masses of the unseen companions are $1.078^{+0.058}_{-0.060}\,M_\odot$ and $0.943^{+0.048}_{-0.049}\,M_\odot$, respectively. Assuming tidal synchronization, we estimate the companion masses to be $1.13^{+0.11}_{-0.08}\,M_\odot$ and $1.05^{+0.15}_{-0.10}\,M_\odot$. The absence of detectable spectral features from the companions rules out main-sequence stars of these masses, suggesting that the unseen companions are likely O/Ne or C/O massive white dwarfs. The short orbital periods imply that these systems are post-common envelope binaries. Their subsequent evolution is uncertain, with possible outcomes including cataclysmic variables, Type Ia supernovae, or accretion-induced collapse, depending on the nature of future mass transfer.

## 1 Introduction

White dwarfs (WDs) in binary systems play a crucial role in the late stages of stellar evolution, producing phenomena such as novae, Type Ia supernovae (SNe), and accretion-induced collapse (e.g., Nomoto et al. 1984; Nomoto & Kondo 1991). In particular, SNe Ia and accretion-induced collapse are closely related to the Chandrasekhar limit of massive WDs. While it is widely accepted that the fate of an accreting WD is determined by its mass, accretion rate, and composition, the specific outcomes for a given set of binary parameters remain uncertain. To better understand their evolutionary pathways, it is therefore important to characterize massive WD binaries and explore the statistical distribution of their masses, orbital periods, and ages across different evolutionary stages (e.g., Zorotovic et al. 2011; Parsons et al. 2016, 2023; Rebassa-Mansergas et al. 2017; Hernandez et al. 2021, 2022a, 2022b).

Recently, large-scale stellar surveys have made it possible to search for dormant single-degenerate binaries (i.e., binaries containing a compact object—WD, neutron star, or black hole—which are currently not accreting mass and hence emitting no X-rays). The astrometric method is effective in identifying such binaries with orbital periods of $\mathcal{O}(1\,\mathrm{yr})$. In fact, black holes and neutron stars are discovered by using the Gaia astrometric solutions (El-Badry et al. 2023a, 2023b, 2024a, 2024b; Chakrabarti et al. 2023; Tanikawa et al. 2023; Gaia Collaboration 2024; Wang et al. 2024). In addition, Yamaguchi et al. (2024c) confirmed about 30 WDs in binaries from the candidates in the Gaia DR3 catalog (Shahaf et al. 2024).

Dormant binaries with shorter periods ($\lesssim$10–100 d) can be identified using high-cadence photometric data and multi-epoch spectroscopic observations. However, finding compact companions next to, for example, low-mass giant stars is challenging, because stellar companions can be easily hidden by giants. Furthermore, if one of the components is a rapidly rotating star, its spectral lines may be significantly broadened or even undetectable. Indeed, most reported close dormant binaries consisting of giants and black-hole candidates (e.g., Thompson et al. 2019; Jayasinghe et al. 2021, 2022) have been challenged (e.g., El-Badry et al. 2022; Bianchi et al. 2024). Such stars are identified as "imposters" [see also, e.g., Abdul-Masih et al. (2020), Bodensteiner et al. (2020), Irrgang et al. (2020), and El-Badry and Quataert (2021) for more imposters].

One way to mitigate these challenges in finding short-period dormant binaries is to focus on those consisting of low-mass main-sequence (MS) stars, where the companions are expected to be brighter if the companions are stars more massive than the primary MS stars. Thus, an MS star orbiting a more massive companion is likely a dormant binary. Neutron-star and WD candidates in close binaries with MS stars have been identified using this method (Faigler et al. 2015; Li et al. 2022; Mazeh et al. 2022; Yi et al. 2022; Yuan et al. 2022; Zheng et al. 2022, 2023; Lin et al. 2023; Qi et al. 2023; Ding et al. 2024; Liu et al. 2024; Rowan et al. 2024; Zhao et al. 2024; Zhang et al. 2024; Tucker et al. 2025; Zhu et al. 2025). Ideally, ellipsoidal variability makes it possible to estimate the orbital inclination and dark companion mass. Although these systems exhibit ellipsoidal variability, their photometric light curves are often significantly contaminated by stellar activities such as spots. As a result, inferring binary parameters using ellipsoidal variation alone is highly non-trivial, and thus, it is challenging to identify dormant binaries using photometric data alone (e.g., Green et al. 2025). Without such estimates on inclination angles, radial velocities alone can only give the minimum masses of the dark companions, which makes it challenging to compare them to binary evolution models.

In this paper, we search for dormant binaries using photometric data from the Transiting Exoplanet Survey Satellite (TESS) and radial velocity variations from Gaia. Combining these datasets significantly reduces the number of false positives due to stellar activities compared to those using photometric data alone. We identify two dormant binaries consisting of unseen companions consistent with massive WDs. In section 2, we introduce the search strategy of identifying dormant binaries. The spectroscopic follow-up observations are described in section 3. In section 4, the stellar and binary parameters are derived. Here the orbital inclinations of the systems are constrained using rotational periods and projected rotational velocities. The possible evolutionary pathways of the two systems are discussed in section 5. We conclude our results in section 6.

## 2 Sample selection

### 2.1 Selection criteria

To search for MS stars orbiting unseen companions in close binaries, we used the amplitudes of radial velocity (RV) variation from the Gaia DR3 catalog and orbital periods inferred from photometric light curves from TESS. The `rv_amplitude_robust` values, which indicate the total amplitude in the RV time series observed by Gaia (Gaia Collaboration 2023a), are used to estimate the semi-amplitudes $K$ of the RVs as (see appendix 1)

$$K \approx \frac{\texttt{rv_amplitude_robust}}{2}. \tag{1}$$

Photometric data from TESS were used to identify periodic variations. We used the single aperture photometry (SAP) light curves generated by the Quick-Look Pipeline group (QLP: Huang et al. 2020a, 2020b; Kunimoto et al. 2021, 2022).[1] The QLP targets have a limiting magnitude of $T < 13.5$, where $T$ is the TESS magnitude. We removed all data points with a non-zero quality flag and all "Not a Number" (NaN) values in light curves.

The photometric period, $P_{\rm LC}$, of each star is determined from the peak of the Lomb–Scargle periodogram of its light curve (Lomb 1976; Scargle 1982). We restricted our samples to stars for which the periods obtained from different TESS sectors are consistent with each other within a 10% level.[2] Systems with $P_{\rm LC} > 10$ d are excluded from our samples. Since variations with such long periods are in principle difficult to confirm by TESS 27.5 d light curves, we consider that periodic variations in these light curves are not detected.

With the two quantities `rv_amplitude_robust` and $P_{\rm LC}$, we constructed a "pseudo-binary mass function":

$$\tilde{f} = \frac{2P_{\rm LC}}{2\pi G}\left(\frac{\texttt{rv_amplitude_robust}}{2}\right)^3, \tag{2}$$

where the orbital period is assumed to be $P_{\rm orb} \simeq 2P_{\rm LC}$. This corresponds to the spectroscopic mass function of circular orbit with orbital period of $2P_{\rm LC}$, and is motivated by the expectation that photometric variations of close binaries are predominately caused by ellipsoidal modulation (e.g., Masuda & Hotokezaka 2019). After removing eclipsing binaries by visual inspection, we selected stars with $\tilde{f} > 0.1M_{\rm MS}^{\rm TIC}$ as binary candidates with massive unseen companions, where $M_{\rm MS}^{\rm TIC}$ is the mass of the visible star provided by the TESS Input Catalog version 8.2 (TIC v8.2: Paegert et al. 2021).

To limit the targets to FGKM-type MS stars, we set constraints on TIC masses $M_{\rm MS}^{\rm TIC}$ and the locations on the color–magnitude diagram as

$$M_{\rm MS}^{\rm TIC} < 1.5\,M_\odot, \tag{3}$$

$$G_{\rm abs} > 5(BP - RP), \tag{4}$$

where $G_{\rm abs}$ is the absolute magnitude in Gaia $G$ band and $BP - RP$ is the color index provided by the Gaia DR3 catalog. Our selection resulted in $\sim 600$ candidates of star–compact object binaries in the northern hemisphere (Dec $> 0$).

Figure 1 shows the color–magnitude diagram of the selected stars along with compact-object imposters reported in the literature. Because of the selection criteria, the stars in our sample lie along the main sequence whereas the imposters are clearly offset from it. Figure 2a presents the distribution of the pseudo-mass functions $\tilde{f}$ against the light curve periods $P_{\rm LC}$ of our sample. We found that about 10% of the stars have $\tilde{f} > 0.5\,M_\odot$, which are good candidates of dormant binaries with massive compact-object companions.

[1] QLP also provides "KSPSAP flux," in which all periodic signals longer than 0.3 d are filtered out. However, we used the un-detrended light curves because the detrending filters out the signals of our interest.

[2] Some objects show different periods $P_{\rm LC}^i$, $P_{\rm LC}^j$ in different sectors $i$ and $j$, where $2P_{\rm LC}^i \simeq P_{\rm LC}^j$. We included these objects in the candidates of MS-compact object binaries if $|2P_{\rm LC}^i - P_{\rm LC}^j| < 0.1P_{\rm LC}^j$. We assumed the orbital period $P_{\rm orb}$ of these objects as $P_{\rm orb} \simeq 2P_{\rm LC}^i$, expecting that photometric variation in sector $i$ is dominated by ellipsoidal variation and variation in sector $j$ by stellar spots.

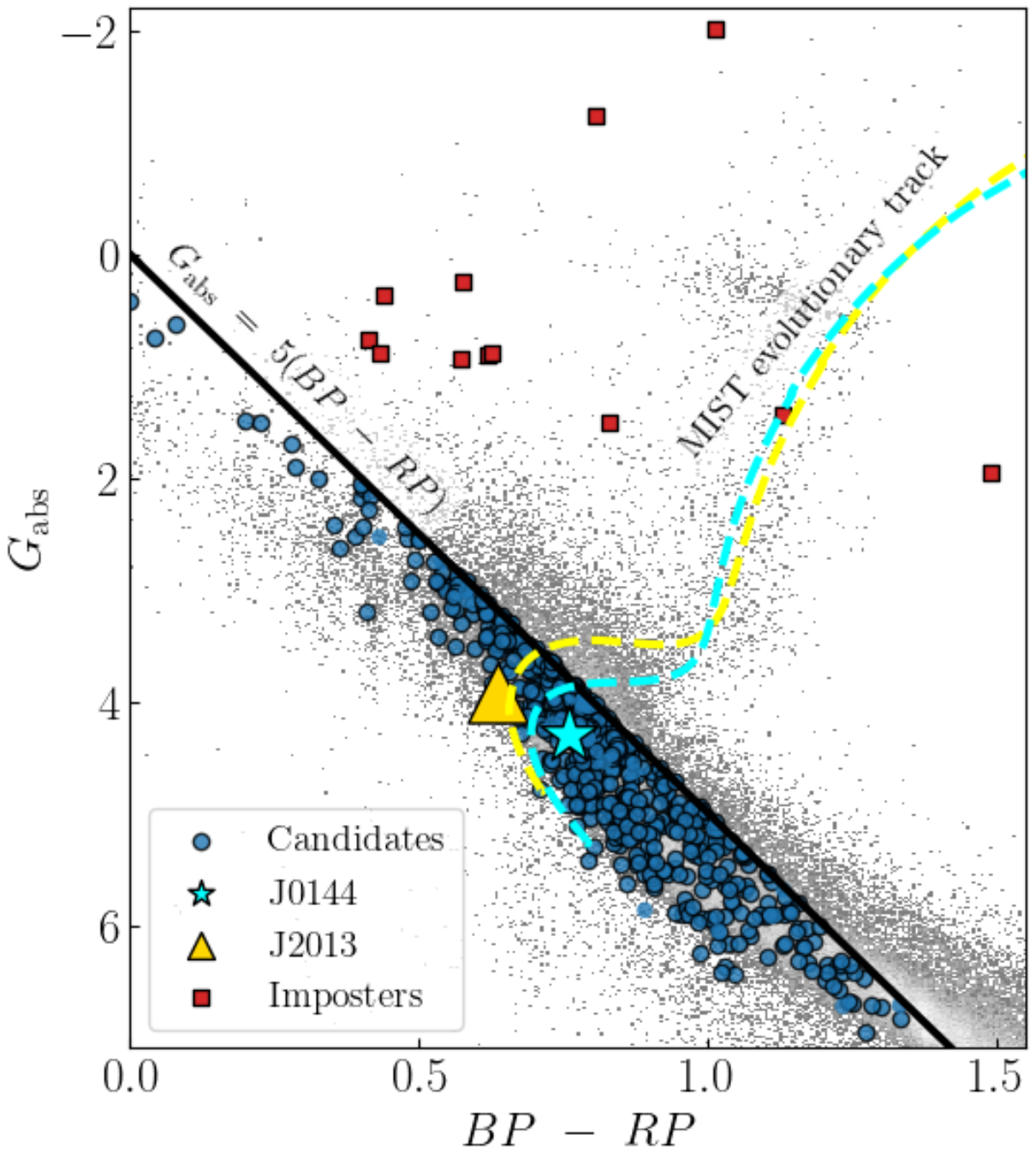

**Fig. 1.** Candidates of star–compact object binaries in the northern hemisphere obtained from our search (circles). The star (J0144) and the triangle (J2013) are the observed targets. Squares are the imposters: Unicorn and Giraffe (El-Badry et al. 2022), and Unicorn-like imposters by El-Badry and Rix (2022). Dashed lines are MIST evolutionary tracks for J0144 (cyan) and J2013 (yellow) based on isochrone fitting (subsection 4.1). On this plot we used extinction correction based on Gaia DR3 `ag_gspphot` and `ebpminrp_gspphot`.

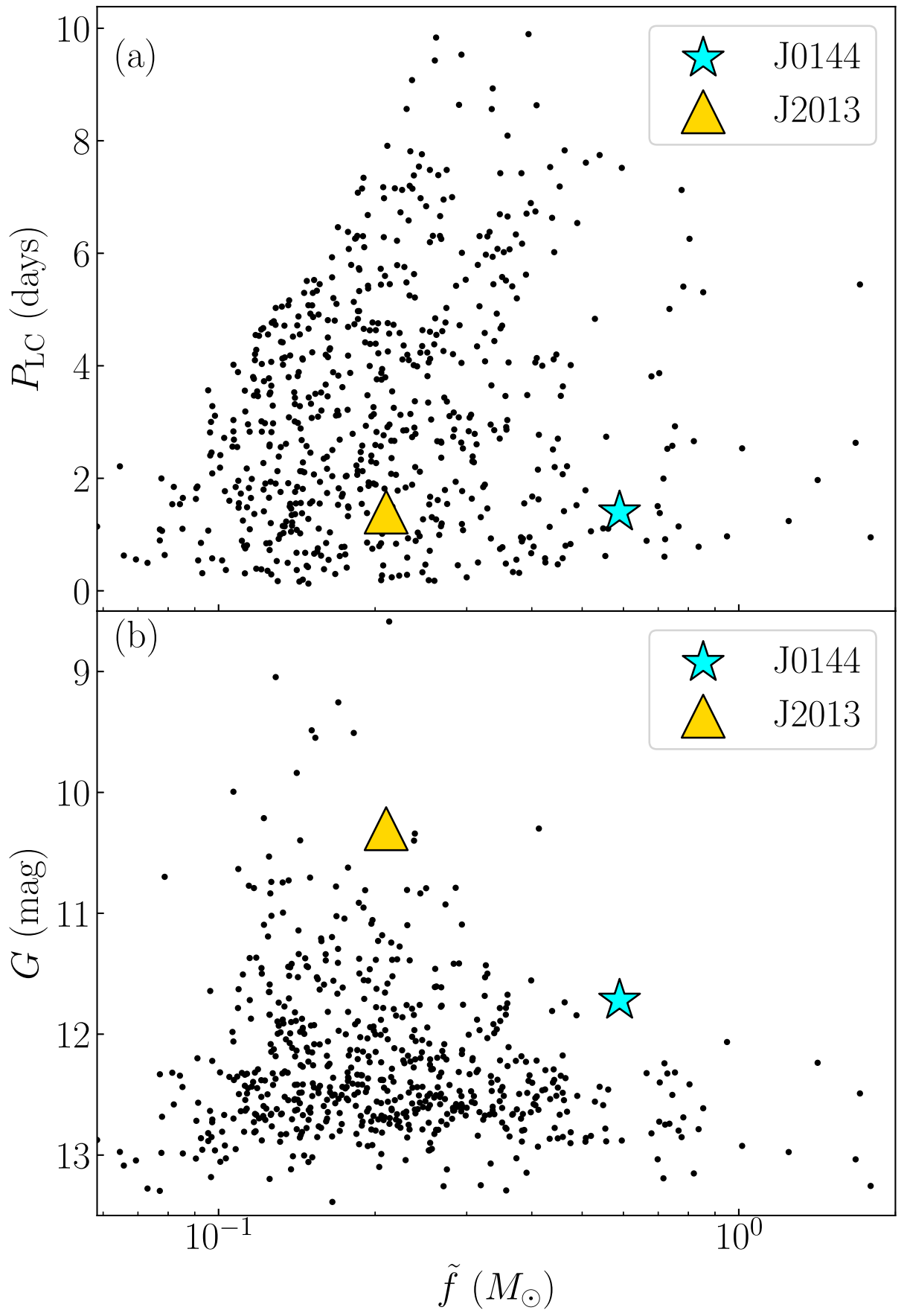

**Fig. 2.** Distribution of pseudo-mass functions $\tilde{f}$ against (a) photometric periods $P_{LC}$ and (b) Gaia *G*-band magnitudes, for the candidates of star–compact object binaries in the northern hemisphere selected according to the criteria in subsection 2.1. The star (J0144) and triangle (J2013) indicate the targets selected for follow-up observations in this study.

## 2.2 Targets in this work

Among the candidates, we conducted spectroscopic follow-up observations of two systems, J0144+5106 and J2013+1734, hereafter J0144 and J2013, respectively. Each of the two observed targets exhibits one of the largest pseudo-mass functions among candidates with comparable *G*-band magnitudes (figure 2b). The parameters of the two objects in the Gaia DR3 catalog and TIC v8.2, along with the derived parameters $P_{LC}$ and $\tilde{f}$, are shown in table 1.

The TESS light curves for these systems are shown in figure 3. The shape of J0144's light curves evolves significantly over time, suggesting that rotational variability caused by stellar spots dominates over ellipsoidal modulation [appendix 4; see also, e.g., Mazeh et al. (2022), Hernandez et al. (2022a, 2022b), and Rowan et al. (2024)]. The light curves of J2013 are more consistent with sinusoidal modulation with a period of $P_{LC} = 1.4$ d, i.e., ellipsoidal variation of $P_{orb} = 2.8$ d. The Lomb–Scargle periodogram (figure 3b, bottom; see also figure 19b) shows another peak at 2.8 d, and the relative amplitudes of the peak at 1.4 d and 2.8 d vary for different TESS sectors. This suggests that the light curve of J2013 is also likely to be affected by stellar spots (see appendix 4).

Note that J2013 is reported as a single-lined spectroscopic binary (SB1) in the Gaia DR3 Non-Single Star (NSS) catalog (Gaia Collaboration 2023b). The orbital solution in the NSS catalog is shown in table 11 in appendix 6. The orbital period reported by the NSS catalog agrees with $2P_{LC} = 2.8$ d.

# 3 Spectroscopic observations and analysis

## 3.1 Observations and data reduction

We observed J0144 and J2013 with Gunma Astronomical Observatory Echelle Spectrograph for Radial Velocimetry (GAOES-RV: Sato et al. 2024) on the 3.8 m Seimei telescope at the Okayama Observatory of Kyoto University. The spectrograph covers 5160–5930 Å, with a spectral resolution of $R \approx 65000$. The typical exposure time was 900–1800 s, with typical signal-to-noise ratios (SNRs) of 20–30 per pixel. The data reduction was done with the `hds_gaoes`[4] package of the `IRAF`[5] software (Tody 1986, 1993; Fitzpatrick et al. 2024), which performs overscan correction, aperture determination, flat fielding, wavelength calibration using ThAr lamps, removal of scattered light, and spectral extraction. The extracted spectra were normalized using the `specutils` package (N. Earl et al. 2021).[6]

Telluric lines were masked using the spectrum of $\alpha$ Aql, an A-type rapid rotator. We calculated the sky transmittance by

[4] ⟨https://github.com/chimari/hds_iraf⟩.
[5] ⟨https://iraf-community.github.io⟩.
[6] N. Earl et al. 2021, astropy/specutils: V1.5.0 ⟨https://zenodo.org/records/5721652⟩.

**Table 1.** Parameters from the Gaia DR3 catalog and TESS.

| | | J0144+5106 | J2013+1734 |
|---|---|---|---|
| | | — Gaia DR3 — | |
| Source ID | | 406025665338103808 | 1809974711191044864 |
| RA* | $\alpha$ (°) | 26.04806436144 | 303.31584864315 |
| Dec* | $\delta$ (°) | 51.11577193221 | 17.56768996582 |
| Parallax | $\varpi$ (mas) | $3.6210 \pm 0.0157$ | $5.3711 \pm 0.0175$ |
| Proper motion in RA | $\mu_{\alpha^*}$ (mas yr$^{-1}$) | $23.321 \pm 0.016$ | $14.067 \pm 0.016$ |
| Proper motion in Dec | $\mu_\delta$ (mas yr$^{-1}$) | $-10.892 \pm 0.015$ | $23.023 \pm 0.012$ |
| Apparent magnitude | $G$ (mag) | 11.72 | 10.30 |
| Color | $BP - RP$ (mag) | 0.88 | 0.65 |
| Distance | $d$ (pc) | $272.4^{+1.7}_{-1.4}$ | $185.04^{+0.56}_{-0.66}$ |
| Radial velocity | `radial_velocity` (km s$^{-1}$) | $-75.18 \pm 26.73$ | $-5.09 \pm 16.74$ |
| RVamp | `rv_amplitude_robust` (km s$^{-1}$) | 255.59 | 180.36 |
| Extinction | $E(BP - RP)$ (mag) | $0.1178^{+0.0043}_{-0.0042}$ | $0.0181^{+0.0033}_{-0.0034}$ |
| Effective temperature | $T_{\rm eff}$ (K) | $5831^{+16}_{-15}$ | $6273^{+14}_{-14}$ |
| Surface gravity | $\log g$ (cgs) | $4.3319^{+0.0038}_{-0.0043}$ | $4.2902^{+0.0062}_{-0.0063}$ |
| Broadening velocity | `vbroad` (km s$^{-1}$) | $86.50 \pm 28.01$ | — |
| Metalicity | [Fe/H] | $-0.319^{+0.016}_{-0.017}$ | $-0.314^{+0.011}_{-0.011}$ |
| | | — TESS — | |
| TIC ID | | 410502040 | 87645385 |
| Primary mass | $M_{\rm MS}^{\rm TIC}$ ($M_\odot$) | $1.10 \pm 0.15$ | $1.27 \pm 0.20$ |
| Primary radius | $R_{\rm MS}^{\rm TIC}$ ($R_\odot$) | $1.105 \pm 0.053$ | $1.154 \pm 0.052$ |
| Primary luminosity | $L_{\rm MS}^{\rm TIC}$ ($L_\odot$) | $1.142 \pm 0.067$ | $1.982 \pm 0.094$ |
| Period of light curve | $P_{\rm LC}$ (d) | 1.4 | 1.4† |
| Pseudo mass function | $\tilde{f}$ ($M_\odot$) | 0.59 | 0.21 |

*Epoch = J2016.0.
†$P_{\rm LC}$ of J2013 is about 2.8 d in sector 54 and 1.4 d in sector 81 (see appendix 4). We estimated that $P_{\rm orb} = 2.8$ d for this system, interpreting that light curve in sector 54 is dominated by spot modulations and that in sector 81 is by ellipsoidal variation.

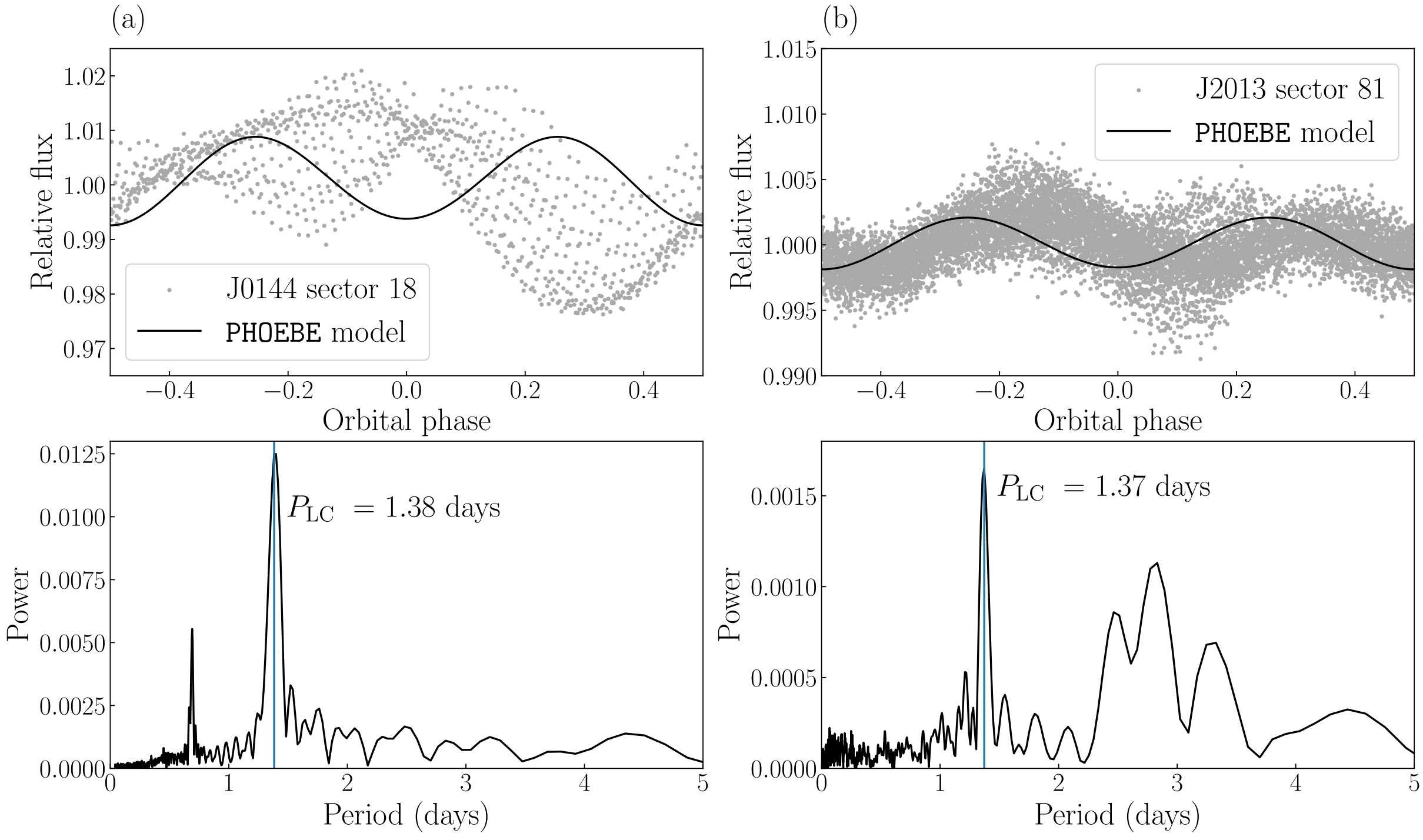

**Fig. 3.** TESS light curve (top) and Lomb–Scargle periodogram (bottom) of (a) J0144 and (b) J2013. Each light curve shows the folded data with the orbital period $P_{\rm orb}$ obtained in section 4. Also depicted are the PHOEBE model light curves of ellipsoidal modulation using the stellar and orbital parameters of J0144 and J2013 obtained in our analysis (appendix 4).

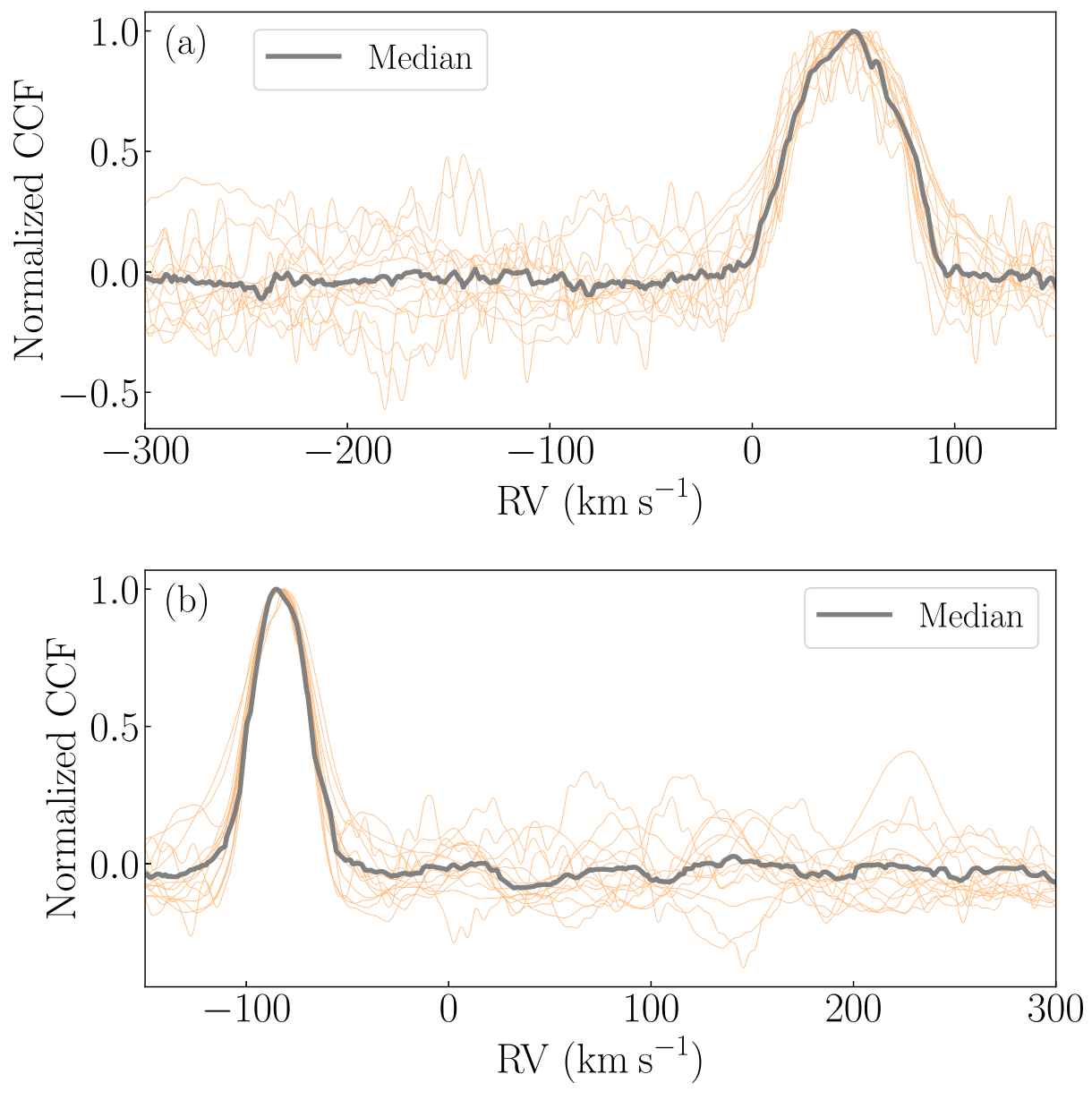

**Fig. 4.** CCF for best SNR spectra of (a) J0144 and (b) J2013. Each orange line is the CCF for each order, and the black line is the median of the CCFs.

dividing the spectrum of $\alpha$ Aql by the median-filtered one with a window width of 101 pixels ($\approx$2 Å). We masked the pixels with a transmittance of less than 98%.

### 3.2 Spectral analysis

The cross-correlation function (CCF: Zucker 2003) of the observed spectra to template spectra was generated to check whether the systems are single-lined or double-lined spectroscopic binaries. Here, the template spectra from the library of synthetic stellar spectra by Coelho et al. (2005) were used.[7] The CCF for the best SNR spectra are shown in figure 4. The CCFs are single-peaked for all taken spectra of both systems, and thus, we conclude that any spectral signatures from a fainter companion are not detected.

To obtain the atmospheric parameters and RVs, we used `jaxspec` (Tomoyoshi et al. 2024),[8] a publicly available python package for spectral fit implemented by `JAX` (J. Bradbury et al. 2018)[9] and `Numpyro` (Bingham et al. 2018; Phan et al. 2019). This package performs linear interpolation of a grid of synthetic spectra, broadening the lines considering macro-turbulence, $v_{\rm rot}\sin i$, and instrumental resolution, and Gaussian process regression for the observed spectrum to derive all the relevant parameters, including RVs. Here we again used the spectral grid by Coelho et al. (2005). Tomoyoshi et al. (2024) take into account the effect of light dilution from the companion, but we ignored it.

The prior distributions in table 1 of Tomoyoshi et al. (2024) were used, except for the projected rotational velocity $v_{\rm rot}\sin i$ and RV. The prior of $v_{\rm rot}\sin i$ was assumed to be uniform in (0, FWHM$_{\rm CCF}$), where FWHM$_{\rm CCF}$ is the full width at half maximum of the CCF of the observed spectra to a template.

The prior of RV was assumed to be uniform in (RV$_{\rm CCF}$ − FWHM$_{\rm CCF}$, RV$_{\rm CCF}$ + FWHM$_{\rm CCF}$), where RV$_{\rm CCF}$ is the RV of the peak of the CCF.

The atmospheric parameters of each star were determined by fitting all 15 orders of the best SNR spectrum. We adopted order-dependent RVs to account for the small offsets in the wavelength calibration, while the other atmospheric parameters are the same for all the orders. The RV of the GAOES-RV spectrum at each epoch was calculated by fitting spectra of orders 7–15 (5144–5620 Å).

We first searched for an optimal set of the parameters that maximizes the likelihood by stochastic variational inferences (SVI: Hoffman et al. 2012), where the posterior was approximated with Gaussian distributions. Here the number of optimization steps and the learning rate were set to be 5000 and 0.01, respectively. Since we found clear data–model mismatches presumably due to missing lines in the model and incomplete masking of cosmic rays, we removed 3 $\sigma$ outliers. Then we ran Hamiltonian Monte Carlo (HMC) sampling to sample the parameter sets from the posterior distribution using the No-U-Turn Sampler (NUTS: Duane et al. 1987; Betancourt 2017) with six chains and 4500 steps. In most cases, the split Gelman–Rubin statistic $\hat{R}$ was $< 1.01$ for each parameter (Gelman et al. 2013), which indicates that HMC most likely converged. The RV, $\bar{v}$, and its statistical uncertainty, $\sigma_{\rm stat}$, at each epoch were evaluated as $\bar{v} = (\sum_i v_i/\sigma_i^2)/\sum_i \sigma_i^{-2}$ and $\sigma_{\rm stat}^{-2} = \sum_i \sigma_i^{-2}$, where $v_i$ and $\sigma_i$ are the mean and standard deviation of RVs in the $i$th order, respectively. Obtained RVs for J0144 and J2013 are shown in tables 3 and 4 in appendix 2, respectively.

The best SNR spectra and the best-fitting models for both systems are shown in figure 5. The estimated atmospheric parameters are summarized in table 2 and the posterior distributions are shown in figures 12 and 13 in appendix 3. The temperatures of the luminous primaries of J1044 and J2013 are, respectively, 5500 and 6000 K, which are consistent with their location on the color–magnitude diagram and with the fact that the primaries are MS stars with TIC mass 1.1 and 1.3 $M_\odot$, respectively.

## 4 Stellar and binary parameters

### 4.1 Isochrone fitting and SED

We derived stellar parameters of the MS stars using `jaxstar`[10] (Masuda 2022), which performs isochrone fitting on the color–magnitude diagram using HMC, implemented by `JAX` and `NumPyro`. The model adopted in this code is based on the MESA Isochrones and Stellar Tracks (MIST) models (Paxton et al. 2011, 2013, 2015; Dotter 2016; Choi et al. 2016) and has three parameters: stellar age $t_\star$, ZAMS metallicity [Fe/H]$_{\rm init}$, and the equivalent evolutionary phase (eep: Dotter 2016). The physical parameters such as effective temperature $T_{\rm eff}$, stellar mass $M_{\rm MS}$, and radius $R_{\rm MS}$ are derived by linearly interpolating model grids for a given set of parameters ($t_\star$, [Fe/H]$_{\rm init}$, eep).

We fitted four observables—$T_{\rm eff}$ and [Fe/H] obtained from GAOES-RV spectral modeling, $K_{\rm s}$-band magnitude $K_{\rm s}$ from the Two Micron All Sky Survey (2MASS: Skrutskie et al. 2006), and parallax $\varpi$ from the Gaia DR3 catalog—by the MIST model. To account for incomplete spectral modeling, we set uncertainty floors of $\Delta T_{\rm eff} = \pm 100$ K and $\Delta$[Fe/H] =

[7] Here we used template with $T_{\rm eff} = 5800$ K, $\log g = 4.4$, [Fe/H] = 0 and [$\alpha$/Fe] = 0. In figure 4, we show CCFs computed by best-fitting spectra, i.e., template with mean parameter values of the posterior of the MCMC by `jaxspec`.

[8] ⟨https://github.com/kemasuda/jaxspec⟩.

[9] J. Bradbury et al. 2018, JAX: composable transformations of Python+NumPy programs ⟨http://github.com/jax-ml/jax⟩.

[10] ⟨https://doi.org/10.5281/zenodo.8272016⟩.

**Table 2.** Estimated parameters for J0144 and J2013.

| | | J0144+5106 | J2013+1734 |
|---|---|---|---|
| | — Atmospheric parameters of G type stars* — | | |
| $\alpha$ element enrichment | [$\alpha$/Fe] | $0.011^{+0.016}_{-0.008}$ | $0.006^{+0.009}_{-0.005}$ |
| Metallicity | [Fe/H] | $-0.26^{+0.06}_{-0.06}$ | $-0.39^{+0.02}_{-0.02}$ |
| Surface gravity | $\log g$ | $3.69^{+0.13}_{-0.13}$ | $4.30^{+0.06}_{-0.06}$ |
| Effective temperature | $T_{\rm eff}$ (K) | $5526^{+80}_{-79}$ | $6083^{+46}_{-44}$ |
| Projected rotational velocity | $v_{\rm rot} \sin i$ (km s$^{-1}$) | $42.86^{+0.96}_{-1.02}$ | $21.16^{+0.57}_{-0.63}$ |
| Macro turbulence velocity | $\zeta$ (km s$^{-1}$) | $8.72^{+0.93}_{-1.72}$ | $5.29^{+0.75}_{-1.01}$ |
| | — Stellar parameters of G type stars† — | | |
| Extinction (fixed) | $A(K)$ | $0.008 \pm 0.002$ | 0 |
| Age | (Gyr) | $12.72^{+0.82}_{-1.58}$ | $7.70^{+1.82}_{-1.70}$ |
| Distance | $d$ (pc) | $276.1^{+1.2}_{-1.2}$ | $186.18^{+0.61}_{-0.63}$ |
| Equivalent evolutionary phase | eep | $424.4^{+2.9}_{-6.2}$ | $420.2^{+9.5}_{-12.6}$ |
| ZAMS metallicity | [Fe/H]$_{\rm init}$ | $-0.067^{+0.077}_{-0.075}$ | $-0.242^{+0.084}_{-0.087}$ |
| Photospheric metallicity | [Fe/H] | $-0.150^{+0.090}_{-0.086}$ | $-0.368^{+0.099}_{-0.099}$ |
| Primary mass | $M_{\rm MS}$ ($M_\odot$) | $0.891^{+0.027}_{-0.020}$ | $0.974^{+0.052}_{-0.050}$ |
| Primary radius | $R_{\rm MS}$ ($R_\odot$) | $1.144^{+0.013}_{-0.013}$ | $1.270^{+0.018}_{-0.018}$ |
| Effective temperature | $T_{\rm eff}$ (K) | $5681^{+66}_{-61}$ | $6093^{+95}_{-98}$ |
| Surface gravity | $\log g$ | $4.271^{+0.018}_{-0.012}$ | $4.219^{+0.031}_{-0.030}$ |
| | — Orbital parameters‡ — | | |
| Orbital period | $P_{\rm orb}$ (d) | $1.370714^{+0.000090}_{-0.000089}$ | $2.75736^{+0.00012}_{-0.00012}$ |
| RV semi-amplitude | $K$ (km s$^{-1}$) | $131.2^{+1.3}_{-1.2}$ | $92.66^{+0.19}_{-0.20}$ |
| Eccentricity | $e$ | 0 (fixed) | 0 (fixed) |
| Center of mass velocity | $\gamma$ (km s$^{-1}$) | $-6.41^{+0.87}_{-0.89}$ | $-30.84^{+0.14}_{-0.16}$ |
| Jitter | $s_{\rm jit}$ (km s$^{-1}$) | $1.90^{+0.71}_{-0.46}$ | $0.49^{+0.16}_{-0.12}$ |
| | — Masses of unseen companions and related parameters§ — | | |
| Mass function | $f$ ($M_\odot$) | $0.3203^{+0.0088}_{-0.0087}$ | $0.2270^{+0.0015}_{-0.0015}$ |
| Minimum companion mass | $M_{\rm c,min}$ ($M_\odot$) | $1.078^{+0.058}_{-0.060}$ | $0.943^{+0.048}_{-0.049}$ |
| Rotational velocity of G-type stars | $v_{\rm rot}$ (km s$^{-1}$) | $44.0^{+2.2}_{-1.4}$ | $22.8^{+1.8}_{-1.2}$ |
| Orbital inclination | $\sin i$ | $0.978^{+0.020}_{-0.051}$ | $0.933^{+0.056}_{-0.075}$ |
| Companion Mass | $M_{\rm c}$ ($M_\odot$) | $1.13^{+0.11}_{-0.08}$ | $1.05^{+0.15}_{-0.10}$ |

*Derived by `jaxspec` spectra fitting of best SNR spectra (section 3).
†Derived by `jaxstar` isochrone fitting (subsection 4.1).
‡Derived by RV fitting (subsection 4.2).
§Derived from the parameters above assuming tidal synchronization (subsection 4.3).

$\pm 0.1$ for effective temperature and [Fe/H], respectively. The dust extinction $A_K$ and its uncertainty $\sigma_{A(K)}$ were evaluated with the `bayestar19` 3D dust map (Green et al. 2019) and the reddening vector derived by Schlafly et al. (2016), based on the Gaia DR3 distance. Instead of fitting the extinction, $\sigma_{A(K)}$ was added to the uncertainty of the $K_{\rm s}$-band magnitude, $\sigma_{K,\rm 2MASS}$: $\sigma_{K,\rm total} = \sqrt{(\sigma^2_{K,\rm 2MASS} + \sigma^2_{A(K)})}$. We did not use magnitudes in other photometric bands following Fulton and Petigura (2018) and Masuda (2022); they do not provide independent constraints given the spectroscopic $T_{\rm eff}$ and can be very sensitive to interstellar extinction.[11] With the prior dis-tribution adopted by Masuda (2022), we drew 5000 samples, after which the resulting chain had the split Gelman–Rubin statistic $\hat{R} < 1.02$ for all the parameters.

[11] For consistency check, we also conducted `jaxstar` isochrone fits using Gaia photometries, and confirmed that the results are reasonably consistent to those without Gaia photometries.

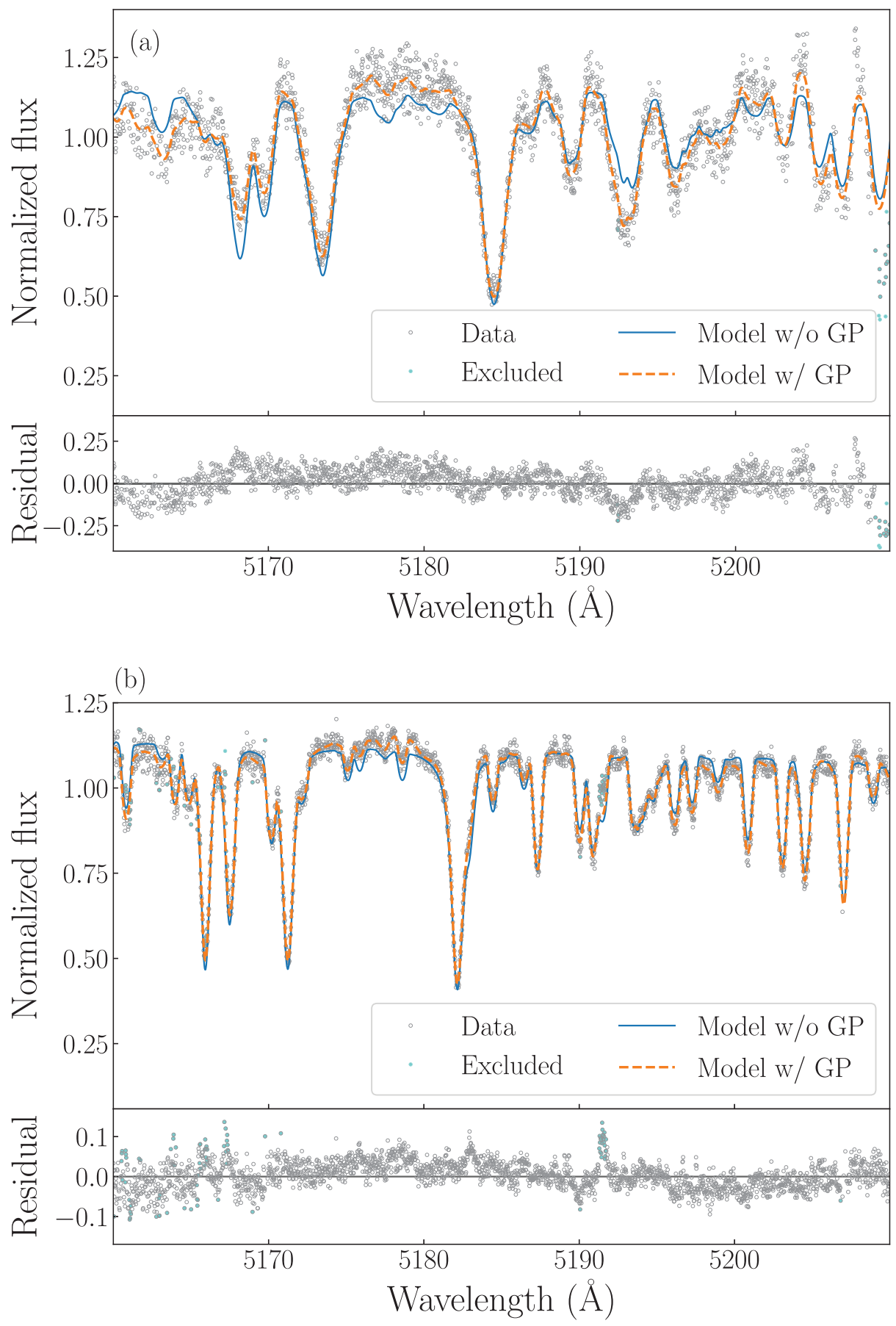

**Fig. 5.** Best-fitting model for best SNR spectrum for (a) J0144 and (b) J2013. The gray dots show data. The light blue dots show the points removed as outliers by SVI. The blue solid lines show a physical spectrum model, computed as the mean of the posterior models. The orange dashed lines show the mean prediction of a model including a Gaussian process noise model.

The results are summarized in table 2, and the posterior distributions of stellar parameters are shown in figures 14 and 15 in appendix 3. The model spectral energy distributions (SEDs) of the best-fitting stellar parameters are compared with the photometric observations in figure 6. We found that the luminous primaries are G-type MS stars with masses of about 0.9–1.0 $M_\odot$, slightly bloated compared with the zero-age main sequence radius.

## 4.2 Radial velocity fitting

We fitted the observed RVs with a circular-orbit model. The likelihood function is computed by

$$\mathcal{L} = \prod_j \frac{1}{\sqrt{2\pi(\sigma_{\mathrm{RV},j}^2 + s_{\mathrm{jit}}^2)}} \exp\left[-\frac{(v_{\mathrm{data},j} - v_{\mathrm{model},j})^2}{2(\sigma_{\mathrm{RV},j}^2 + s_{\mathrm{jit}}^2)}\right], \quad (5)$$

where $v_{\mathrm{data},j}$ and $v_{\mathrm{model},j}$ are the RV data and model at the $j$th epoch, respectively; $\sigma_{\mathrm{RV},j}$ is the RV uncertainty at the $j$th epoch; $s_{\mathrm{jit}}$ is the RV jitter that takes into account any additional scatter in the data that is not included in the model. The prior distributions of the RV semi-amplitude $K$, the orbital period $P_{\mathrm{orb}}$, the initial phase $\theta_0$ at the first RV epoch, and the RV $\gamma$ of the center of mass of the system are assumed to be uniform in $0 < K < 200\,\mathrm{km\,s^{-1}}$, $0.1 < P_{\mathrm{orb}} < 3.0\,\mathrm{d}$, $-0.5 < \theta_0/2\pi < 0.5$, and $-40 < \gamma < 40\,\mathrm{km\,s^{-1}}$, respectively. The prior of the RV jitter $s_{\mathrm{jit}}$ is assumed to be log-uniform in $10^{-5} < s_{\mathrm{jit}} < 10^{1.3}\,\mathrm{km\,s^{-1}}$.

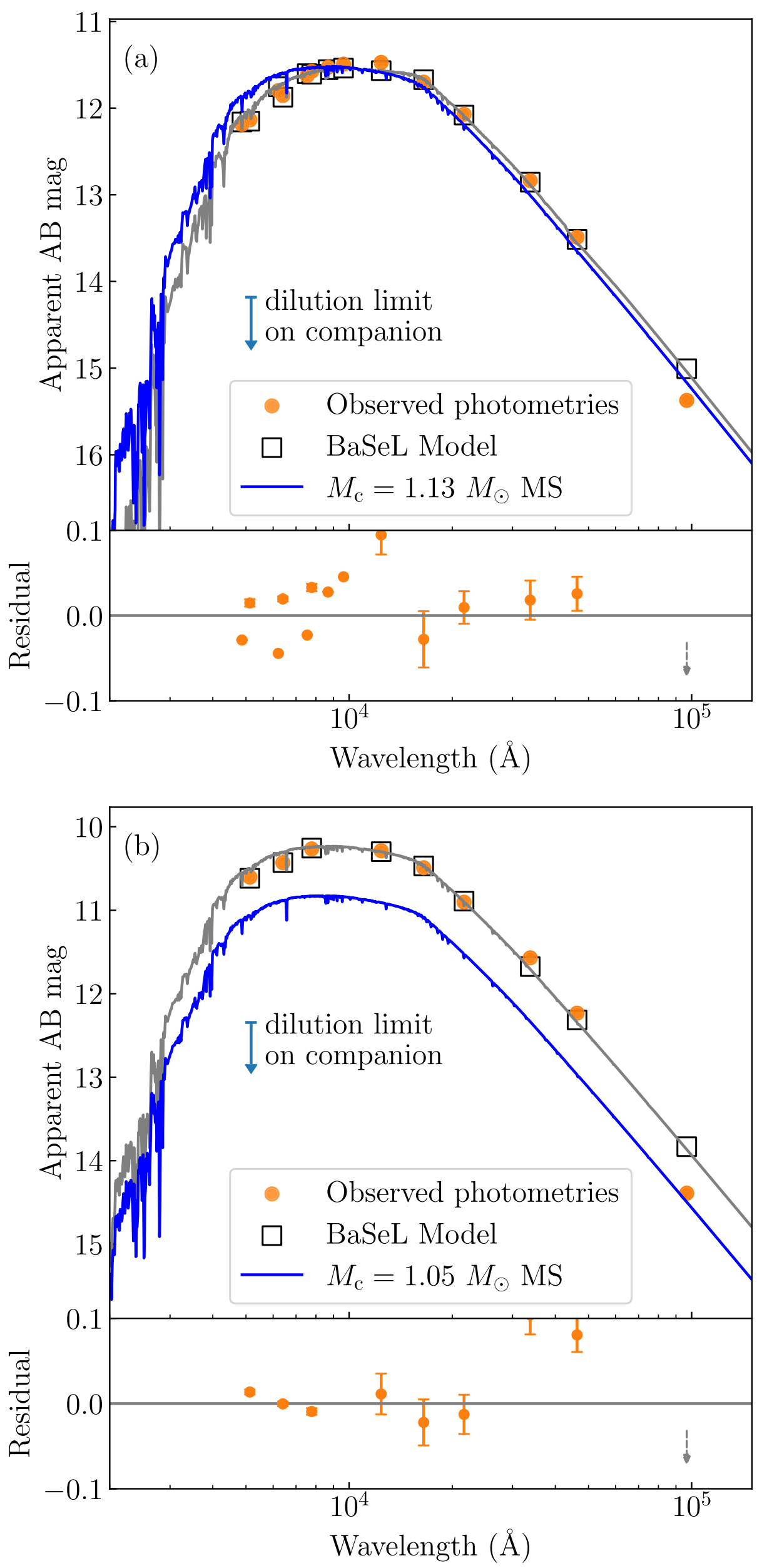

**Fig. 6.** SED for (a) J0144 and (b) J2013. The orange points show the observed photometric data (see tables 9 and 10). The gray lines show the model SED of the luminous primaries based on stellar parameters inferred from `jaxstar` isochrone fitting. The blue curves show the expected SEDs of MS stars with masses equal to the inferred companion mass ($M_c$) for each system. Here the model SEDs are calculated using BaSel library (v2.2: Lejeune et al. 1997, 1998), `pystellibs`,[10] and `pyphot`.[3] Also depicted is the upper limit of flux dilution from unseen companion with 97.5% confidence level ("Dilution limit on companion," see subsection 4.3). In the bottom subfigure of each panel, the gray dashed arrow indicates the direction of the residual point of WISE *W3* band.

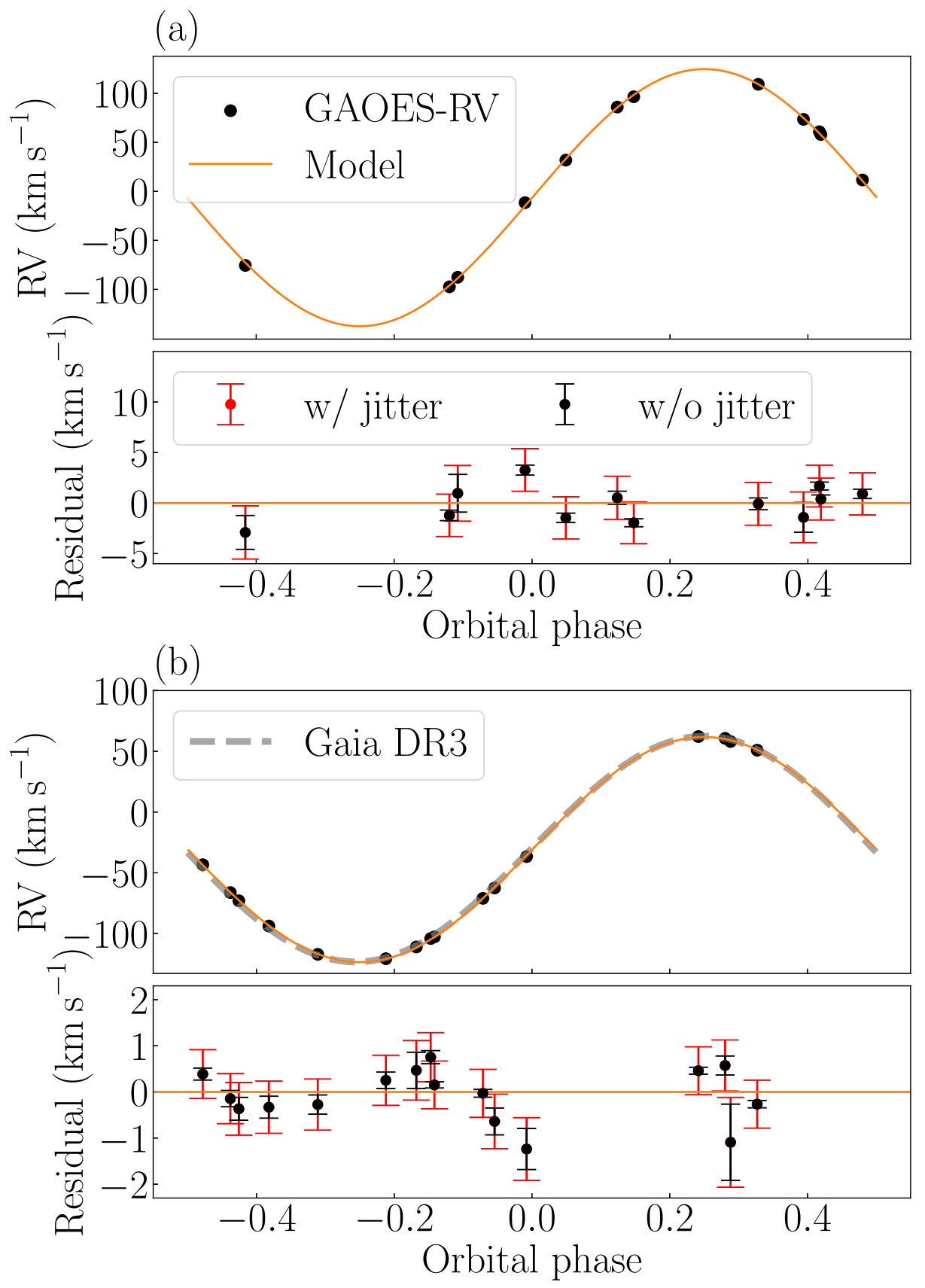

**Fig. 7.** RV curve for (a) J0144 and (b) J2013. At the bottom subfigure of each panel, the black error bars show the statistical uncertainty $\sigma_{\rm stat}$ and the red error bars show the uncertainty $\sigma_{\rm tot}$ that the jitter $s_{\rm jit}$ is taken into account [i.e., $\sigma_{\rm tot} = \sqrt{(\sigma^2_{\rm stat} + s^2_{\rm jit})}$]. In panel (b) the gray dashed curve shows the orbital solution by Gaia DR3 NSS catalog with an arbitrary phase shift.

We used importance nested sampling (Feroz et al. 2019) with pymultinest[12] (Buchner et al. 2014; Buchner 2016), the python interface of the MultiNest package (Feroz & Hobson 2008; Feroz et al. 2009), which allows efficient Bayesian parameter estimation and sampling from multi-modal posterior distribution. We explored the parameter space using 4000 MultiNest live points. The sampling efficiency is set to be 0.1. In each analysis, we enabled Multi-Nest to evaluate multiple modes. To check convergence and consistency of the results, we additionally performed the sampling with different numbers of live points and efficiencies. The results presented here are consistent with the number of live points varying from 4000 to 12000 and an efficiency of 0.1–0.01 [see Dittmann (2024) for a detailed discussion on MultiNest's hyperparameters].

The results are summarized in table 2. The RV curves of the best-fitting model are shown in figure 7, where orbital phase is defined such that the RV maximum occurs at phase $\phi = 0.25$. The posterior distributions of orbital parameters are shown in figures 16 and 17 in appendix 3. For both systems, we obtained RV amplitudes consistent with rv_amplitude_robust from the Gaia DR3 catalog. The orbital period $P_{\rm orb}$ obtained by the RV fit is $\approx 2P_{\rm LC}$ for J2013 as in our expectation, while $P_{\rm orb} \approx P_{\rm LC}$ for J0144. The binary parameters obtained for J2013 are consistent with those in the Gaia NSS catalog.

We obtained the jitter $s_{\rm jit}$, additional uncertainties for RVs, similar to or larger than the statistical uncertainties of RVs as shown in figure 7. The systems may also show some phase-dependent variations in RV residuals whose amplitudes are similar to the jitter. We consider that one possible cause of these is the effect of stellar spots. Aigrain et al. (2012) estimated the amplitude $\Delta{\rm RV}_{\rm rot}$ of RV modulation due to spots as

$$\Delta{\rm RV}_{\rm rot} = \frac{\delta F}{F} v_{\rm rot} \sin\, i \cos\, \delta$$
$$= 0.5\,{\rm km\,s^{-1}} \left(\frac{\delta F/F}{1\%}\right)\left(\frac{v_{\rm rot}\sin\, i}{50\,{\rm km\,s^{-1}}}\right) \cos\, \delta, \quad (6)$$

where $\delta F/F$ is the relative drop in flux due to the spot and $\delta$ is the latitude of the spot relative to the star's rotational equator. We obtained $\Delta{\rm RV}_{\rm rot} \approx (2\text{–}4\,{\rm km\,s^{-1}}) \cos\delta$ for J0144 and $(0.3\text{–}0.4\,{\rm km\,s^{-1}}) \cos\delta$ for J2013.[13] These are close to the value of the RV jitter and the modulation of RV residuals relative to the circular orbit model.

## 4.3 Masses of the unseen companions

Based on the orbital parameters, the minimum companion mass $M_{\rm c,min}$ is obtained by setting $\sin\, i = 1$, leading to $M_{\rm c,min} = 1.078^{+0.058}_{-0.060}\,M_\odot$ and $0.943^{+0.048}_{-0.049}\,M_\odot$ for J0144 and J2013, respectively. For both systems, the minimum mass ratio $q_{\rm min} \equiv (M_{\rm c}/M_{\rm MS})_{\rm min} \gtrsim 1$ implies that the unseen companions are compact objects.

To constrain $\sin\, i$, we assumed tidal synchronization between the stellar rotation of the luminous stellar components and the orbit. By using the orbital period $P_{\rm orb}$, stellar radius $R_{\rm MS}$ derived by the isochrone fit, and the projected rotational velocity $(v_{\rm rot} \sin\, i)_{\rm spec}$ from the spectra, $\sin\, i$ is evaluated as

$$\sin\, i \approx \frac{P_{\rm orb}(v_{\rm rot}\sin\, i)_{\rm spec}}{R_{\rm MS}}. \quad (7)$$

The posterior distribution of the companion mass was calculated by HMC with 100000 NUTS samples. The likelihood function is computed as

$$\mathcal{L} = \frac{1}{\sqrt{2\pi}\sigma_{v\sin\, i}} \times \exp\left\{-\frac{\left[(v_{\rm rot}\sin\, i)_{\rm spec} - (v_{\rm rot}\sin\, i)_{\rm model}\right]^2}{2\sigma^2_{v\sin\, i}}\right\}, \quad (8)$$

where $(v_{\rm rot}\sin\, i)_{\rm model} = R_{\rm MS}\sin\, i/P_{\rm orb}$. We assumed the prior distributions of $K$, $M_{\rm MS}$, and $R_{\rm MS}$ to be normal distributions with means and standard deviations obtained by orbital, isochrone, and spectral fit, and the uniform prior for $\cos\, i$. We set the uncertainty floor of $0.1\,M_\odot$ and $0.1\,R_\odot$ for $M_{\rm MS}$ and $R_{\rm MS}$, respectively, to account for the imperfectness of the MIST model. The orbital period was fixed to the mean value obtained by the RV fit because the uncertainty is very small. The results are summarized in table 2, and the posterior distributions of the parameters are shown in figures 8 and 9.

The estimated masses of the unseen companions in both systems lie between 1.0 and 1.2 $M_\odot$. Although the upper ends of these masses are around the lower edge of the neutron-star mass distribution (e.g., Kiziltan et al. 2013), these unseen companions are more likely massive WDs.

[12] ⟨https://github.com/JohannesBuchner/PyMultiNest⟩.

[13] Here $\delta F/F$ is calculated as the difference between the minimum and maximum of the residuals of the TESS light curve from the PHOEBE ellipsoidal model (see appendix 4).

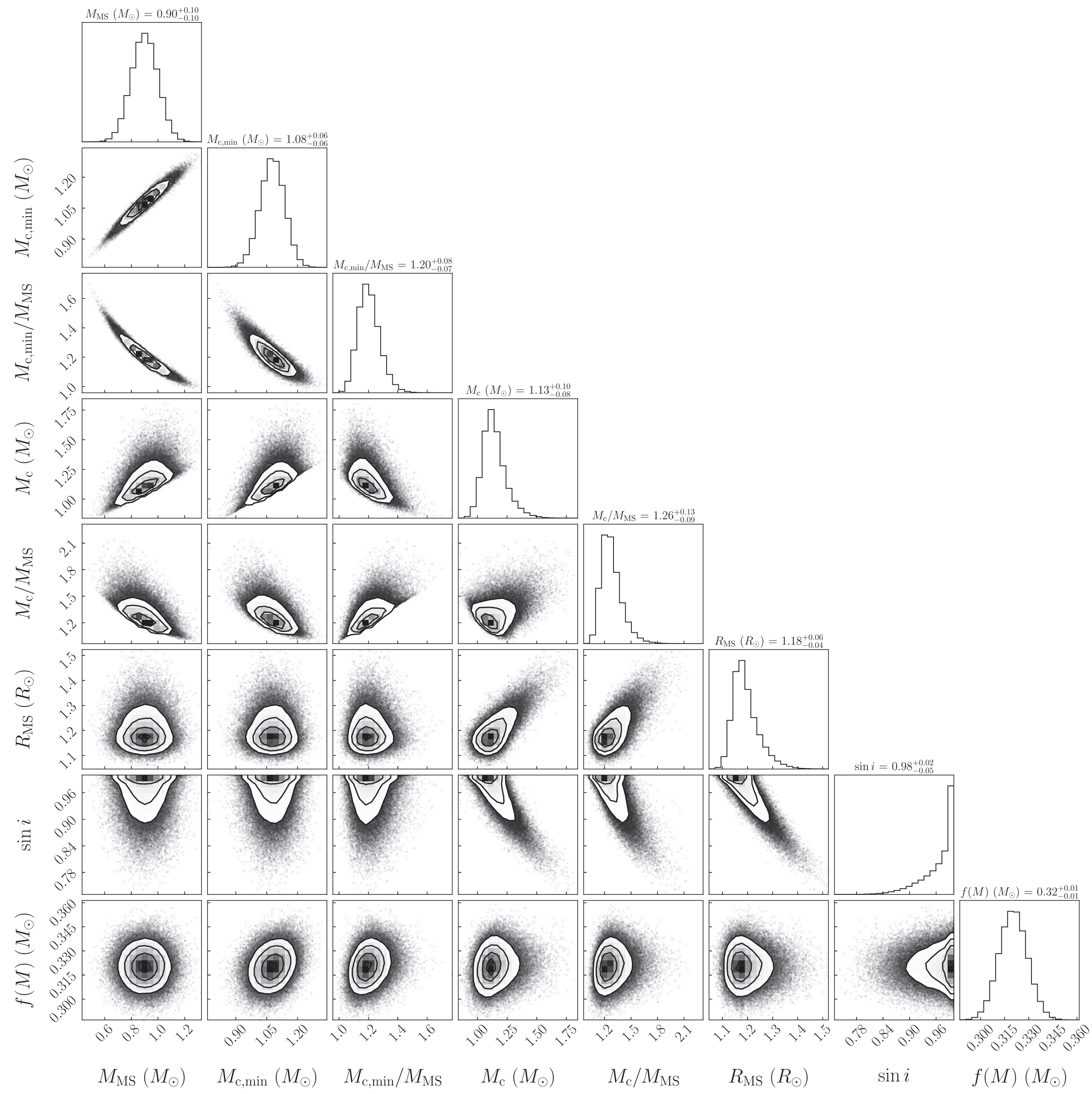

**Fig. 8.** Posterior distribution of companion mass and related parameters for J0144, assuming tidal synchronization (subsection 4.3).

Finally, in order to test the possibility that the unseen companions are MS stars, we fitted the best SNR spectra accounting for the light dilution from the unseen companions and calculated the possible maximum fraction of dilution, following Tomoyoshi et al. (2024). We obtained the fraction of dilution consistent with zero, and the upper limit of dilution fraction with 97.5% statistical significance for J0144 and J2013 are 15% and 19%, respectively. The blue lines in figure 6 show the model SEDs of MS stars with the estimated mass of the unseen companions. Since these SED models exceed the dilution limits, we conclude that the luminous primaries cannot outshine the companions if they are MS stars.

## 5 Discussion on the evolutionary pathways

Figure 10 shows the orbital periods and WD masses of known dormant WD-MS binaries, including J0144 and J2013. These two systems lie at the upper right of the distribution of post-common envelope binaries with short ($\lesssim$10 d) orbital periods. The MS stars are relatively massive compared to typical post-common envelope binaries. These features are consistent with standard common envelope evolution scenarios, where the WD progenitor stars had Zero-Age Main Sequence (ZAMS) masses in the range from 6 to 8 $M_\odot$ and the initial orbits were much wider than the present ones. Using the standard $\alpha\lambda$ prescription with $\alpha\lambda = 1$ and a ZAMS mass of 6 $M_\odot$, the donor radii for J0144 and J2013 are estimated as 400 $R_\odot$ and 600 $R_\odot$, respectively. These values are consistent with radii of stars on the asymptotic giant branch (AGB), suggesting that the common envelope phase was initiated during this phase.

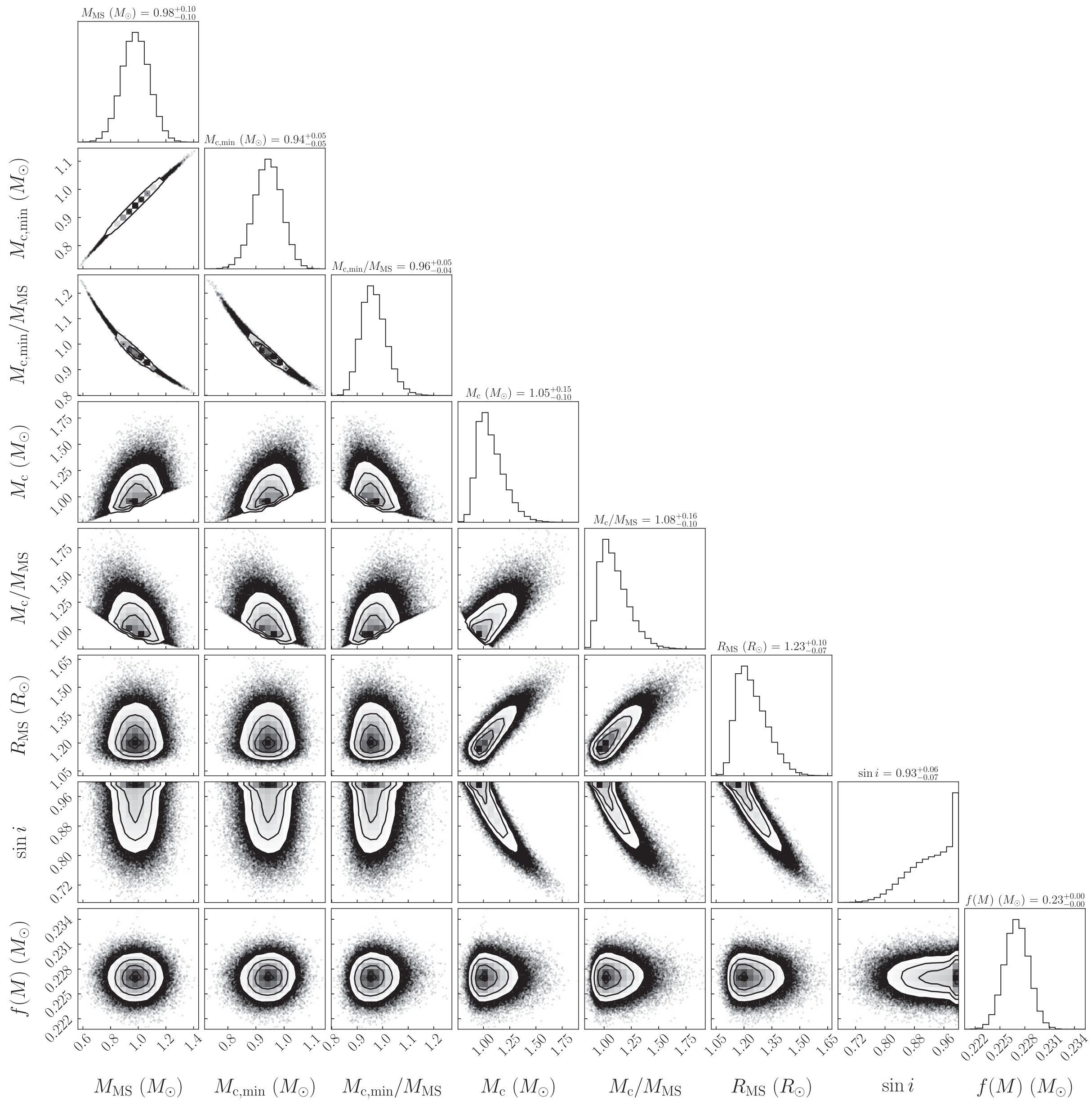

**Fig. 9.** Same as figure 8, but for J2013.

The subsequent evolution of J0144 and J2013 is uncertain because it depends on the stability and timescale of the next mass transfer event. For instance, if the transfer occurs during the Hertzsprung gap phase, the mass transfer is expected to proceed on a thermal time scale. In contrast, if it starts during the red giant phase, the mass transfer may become dynamically unstable, leading to a second common envelope phase.

If the mass transfer proceeds on a thermal timescale, the subsequent evolution depends on the mass transfer rate. Given a thermal timescale of $\sim 10^7$ yr for the donor stars, the expected mass transfer rate is up to $10^{-7}\,M_\odot\,\mathrm{yr}^{-1}$. According to Nomoto and Kondo (1991), if $\dot{M} \gtrsim 10^{-8}\,M_\odot\,\mathrm{yr}^{-1}$, accretion-induced collapse is expected for O/Ne/Mg WDs and an SN Ia may occur for C/O WDs.[14] In the case of the lower accretion rate, $\dot{M} \lesssim 10^{-8}\,M_\odot\,\mathrm{yr}^{-1}$, an off-center helium detonation may be triggered, preventing the WD from growing in mass [see also Nomoto and Leung (2018) for a recent review on the case of C/O WDs].

If the next mass transfer is dynamically unstable, the outcome is expected to be an ultracompact binary consisting of a massive WD and an He WD with an orbital period of $\sim$30 min for $\alpha\lambda = 1$.[15] The orbit further contracts due to gravitational-wave radiation and the mass transfer from the He WD occurs. If this mass transfer is stable, an AM CVn system may form (e.g., Paczyński 1967; Kilic et al. 2014). On the contrary, if the last mass transfer is unstable, the two WDs merge (e.g., Marsh et al. 2004; Shen 2015), which results in various possible outcomes, including the formation of an R CrB/Hydrogen-deficient carbon star (e.g., Webbink 1984; Iben et al. 1996; Clayton 2012), a normal or peculiar SN Ia (e.g., Perets et al. 2010; Shen 2015; Ferrario 2025), or accretion-induced collapse (e.g., Nomoto & Kondo 1991).

[14] Accretion-induced collapse may be prevented by mass loss by wind associated with shell flushes (e.g., Kato & Hachisu 1989).

[15] Two WDs are expected to merge if $\alpha\lambda \lesssim 0.1$.

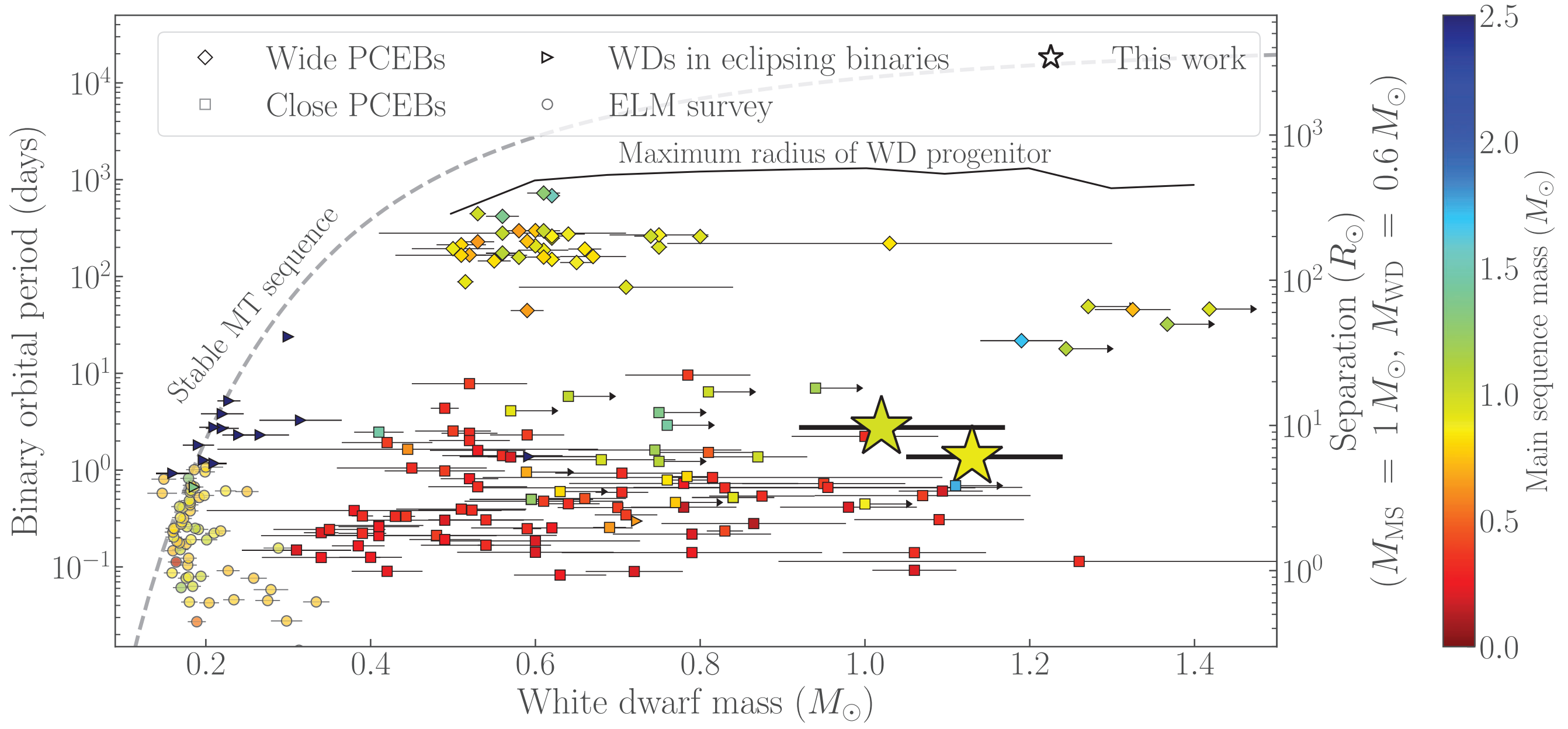

**Fig. 10.** Masses of WDs in binaries and their orbital periods. On the right-hand axis, we show the minimum orbital separation corresponding to the period assuming $M_{\rm MS} = 1\,M_\odot$ and $M_{\rm WD} = 0.6\,M_\odot$. The marker colors represent the mass of the MS stars in the binary systems. Star marks: the systems discovered in this work (J0144 and J2013). Diamonds: literature post common envelope binaries with $P_{\rm orb} \gtrsim 10$ d [IK Peg (Landsman et al. 1993; Wonnacott et al. 1993; Davis et al. 2010), Kepler self-lensing binaries (Kruse & Agol 2014; Kawahara et al. 2018; Masuda et al. 2019; Yamaguchi et al. 2024b), WD binary pathways survey (Parsons et al. 2023), Gaia DR3 NSS Astrometry (Yamaguchi et al. 2024c), and Gaia DR3 NSS SB1 (Yamaguchi et al. 2024a)]. Squares: literature post common envelope binaries with $P_{\rm orb} \lesssim 10$ d [V471 Tau (O'Brien et al. 2001), SDSS (Rebassa-Mansergas et al. 2010, 2016), TYC 6760-497-1 (Parsons et al. 2015), WD binary pathways survey (Hernandez et al. 2021, 2022a, 2022b), and LAMOST (Li et al. 2022; Zheng et al. 2022; Qi et al. 2023; Ding et al. 2024; Rowan et al. 2024; Zhang et al. 2024; Zhu et al. 2025)]. Triangles: WDs in eclipsing binaries [Kepler (van Kerkwijk et al. 2010; Carter et al. 2011; Breton et al. 2012; Muirhead et al. 2013; Faigler et al. 2015; Rappaport et al. 2015; Zhang et al. 2017), OGLE and WASP surveys (Pietrzyński et al. 2012; Maxted et al. 2013), and GPX-TF 16E–48 (Krushinsky et al. 2020)]. Circles: binaries with extremely low mass WDs by Brown et al. (2016). The gray dashed line is the theoretical $P$-$M_{\rm WD}$ relation for the stable mass transfer case by Lin et al. (2011). The black solid line is the maximum radius that is reached by the giant progenitor at each WD mass by Yamaguchi et al. (2024c).

## 6 Conclusion

We searched for "dormant" close binaries consisting of MS stars and unseen compact companions using photometric data from TESS and RV amplitudes from Gaia, resulting in approximately 600 candidates in the northern hemisphere. Among them, we conducted high-resolution spectroscopic observations for several objects and identified two dormant binaries, J0144 and J2013, with orbital periods of 1.37 and 2.76 d, respectively.

The visible components of J0144 and J2013 are slightly bloated G-type MS stars with masses of $\sim$0.9–1.0 $M_\odot$. The minimum masses of the unseen objects are $1.078^{+0.058}_{-0.060}\,M_\odot$ for J0144 and $0.943^{+0.048}_{-0.049}\,M_\odot$ for J2013. Assuming tidal synchronization, their masses are estimated to be $1.13^{+0.11}_{-0.08}\,M_\odot$ and $1.05^{+0.15}_{-0.10}\,M_\odot$, respectively. We found that the photometric light curves are significantly affected by stellar spots, and thus, it is difficult to estimate binary parameters using ellipsoidal variation. The absence of spectral features of the subluminous companions rules out MS stars with the inferred masses, suggesting the unseen companions are likely O/Ne or C/O massive WDs.

Given their short orbital periods, both J0144 and J2013 are consistent with being post-common envelope binaries. These systems likely underwent a common envelope phase when the WD progenitors were in the AGB phase. Using the standard common envelope prescription with $\alpha\lambda = 1$ and a ZAMS mass of 6 $M_\odot$, the donor radii at the onset of the common envelope phase are estimated as 400 $R_\odot$ for J0144 and 600 $R_\odot$ for J2013.

Their subsequent evolution is rather uncertain. As the visible components evolve and initiate mass transfer, the outcomes will depend on the stability of the mass transfer. If the transfer is dynamically stable, the systems may evolve into CVs, undergo accretion-induced collapse, or trigger SNe Ia, depending on the accretion rates (Nomoto & Kondo 1991). If dynamically unstable, the systems are expected to become binaries consisting of a low-mass He WD and a massive WD. Such systems will further evolve into AM CVn systems, R CrB stars, normal/peculiar SNe Ia, or accretion-induced collapse.

Our results demonstrate that searching for close binaries composed of MS stars and unseen companions more massive than the visible components, using data from TESS and Gaia, is an effective approach. The discovery of J0144 and J2013 suggests many more such systems remain to be identified in existing and upcoming datasets. Revealing the statistical distribution of massive WDs orbiting around MS stars will provide important insights into the diverse evolutionary pathways of these binary systems.

## Data availability

TESS full-frame images are available from the Barbara A. Mikulski Archive for Space Telescopes (MAST),[16] as are the QLP extracted lightcurves.[17] The Gaia data can be retrieved through the Gaia archive.[18] The GAOES-RV spectra can be shared on reasonable request to the corresponding author, and will be publicly available through SMOKA Science Archive.[19] Any other data can be made available upon reasonable request to the corresponding author.

## Funding

KH is supported by the Japan Science and Technology Agency (JST) Fusion Oriented REsearch for disruptive Science and Technology (FOREST) Program (JPMJFR2136) and the Japan Society for the Promotion of Science (JSPS) Grant-in-Aid for Scientific Research (20H05639, 20H00158, 23H01169, 23H04900). KM is supported by JSPS Grant-in-Aid for Scientific Research (25K07387). SH is supported by Grants-in-Aid for Scientific Research (22K03688). AT is supported by Grants-in-Aid for Scientific Research (24K07040). TT is supported by Grants-in-Aid for Scientific Research (24KJ0605).

## Acknowledgments

We thank Daniel Kasen, Tomoya Kinugawa, and Ken'ichi Nomoto for helpful conversations. We would like to express our gratitude to all the staff members at Nishi-Harima Astronomical Observatory and Okayama Observatory for their support during observations. We also acknowledge the dedicated support of Fumihiro Naokawa during observations.

YS is supported by Forefront Physics and Mathematics Program to Drive Transformation (FoPM), a World-leading Innovative Graduate Study (WINGS) Program, the University of Tokyo. SJ acknowledges support from the Japan Society for the Promotion of Science (JSPS) as a postdoctoral fellow.

The GAOES-RV project started as a collaboration between Gunma Astronomical Observatory and Institute of Science Tokyo. Based on the contract signed between the two parties, GAOES-RV is lent to Institute of Science Tokyo and operated at the Seimei Telescope.

This work has made use of data from the European Space Agency (ESA) mission Gaia,[20] processed by the Gaia Data Processing and Analysis Consortium (DPAC.).[21] Funding for the DPAC has been provided by national institutions, in particular the institutions participating in the Gaia Multilateral Agreement.

We acknowledge the use of TESS High Level Science Products (HLSP) produced by the Quick-Look Pipeline (QLP) at the TESS Science Office at MIT, which are publicly available from the Mikulski Archive for Space Telescopes (MAST) at the Space Telescope Science Institute (STScI). Funding for the TESS mission is provided by NASA's Science Mission directorate. STScI is operated by the Association of Universities for Research in Astronomy, Inc., under NASA contract NAS 5-26555.

16 ⟨https://archive.stsci.edu/missions-and-data/tess⟩.
17 ⟨https://archive.stsci.edu/hlsp/qlp⟩.
18 ⟨https://gea.esac.esa.int/archive⟩.
19 ⟨https://smoka.nao.ac.jp⟩.
20 ⟨https://www.cosmos.esa.int/gaia⟩.
21 ⟨https://www.cosmos.esa.int/web/gaia/dpac/consortium⟩.

This publication has made use of data products from the Two Micron All Sky Survey, which is a joint project of the University of Massachusetts and the Infrared Processing and Analysis Center/California Institute of Technology, funded by the National Aeronautics and Space Administration and the National Science Foundation.

The Pan-STARRS1 Surveys (PS1) and the PS1 public science archive have been made possible through contributions by the Institute for Astronomy, the University of Hawaii, the Pan-STARRS Project Office, the Max-Planck Society and its participating institutes, the Max Planck Institute for Astronomy, Heidelberg and the Max Planck Institute for Extraterrestrial Physics, Garching, The Johns Hopkins University, Durham University, the University of Edinburgh, the Queen's University Belfast, the Harvard-Smithsonian Center for Astrophysics, the Las Cumbres Observatory Global Telescope Network Incorporated, the National Central University of Taiwan, the Space Telescope Science Institute, the National Aeronautics and Space Administration under Grant No. NNX08AR22G issued through the Planetary Science Division of the NASA Science Mission Directorate, the National Science Foundation Grant No. AST-1238877, the University of Maryland, Eotvos Lorand University (ELTE), the Los Alamos National Laboratory, and the Gordon and Betty Moore Foundation.

This publication has made use of data products from the Wide-field Infrared Survey Explorer, which is a joint project of the University of California, Los Angeles, and the Jet Propulsion Laboratory/California Institute of Technology, funded by the National Aeronautics and Space Administration.

This work has made use of `Astropy`:[22] a community-developed core Python package and an ecosystem of tools and resources for astronomy (Astropy Collaboration 2013, 2018, 2022).

NOIRLab IRAF is distributed by the Community Science and Data Center at NSF NOIRLab, which is managed by the Association of Universities for Research in Astronomy (AURA) under a cooperative agreement with the U.S. National Science Foundation.

This research has made use of the VizieR catalogue access tool, CDS, Strasbourg, France ⟨https://cds.unistra.fr/⟩. The original description of the VizieR service was published in Ochsenbein, Bauer, and Marcout (2000).

## Appendix 1 Radial velocities in the Gaia DR3 and SB9 catalogs

Our search assumes that the values of $K$, the semi-amplitude of RV variation, can be inferred from the `rv_amplitude_robust` parameter in the Gaia DR3 catalog. If the variation is due to binary orbital motion, `rv_amplitude_robust` is expected to approximate $2K$. To validate this assumption, we compare `rv_amplitude_robust` with the known amplitudes of spectroscopic binaries from the SB9 catalog (Pourbaix et al. 2004). We obtained a cross-matched sample of 1251 objects whose sky positions agree within 1″. Figure 11 compares the `rv_amplitude_robust` values with $2K$ from SB9. For the majority of objects, `rv_amplitude_robust` agrees with $2K$ within $\sim$10%. In cases where the two values differ significantly, `rv_amplitude_robust` tends to underestimate

22 ⟨http://www.astropy.org⟩.

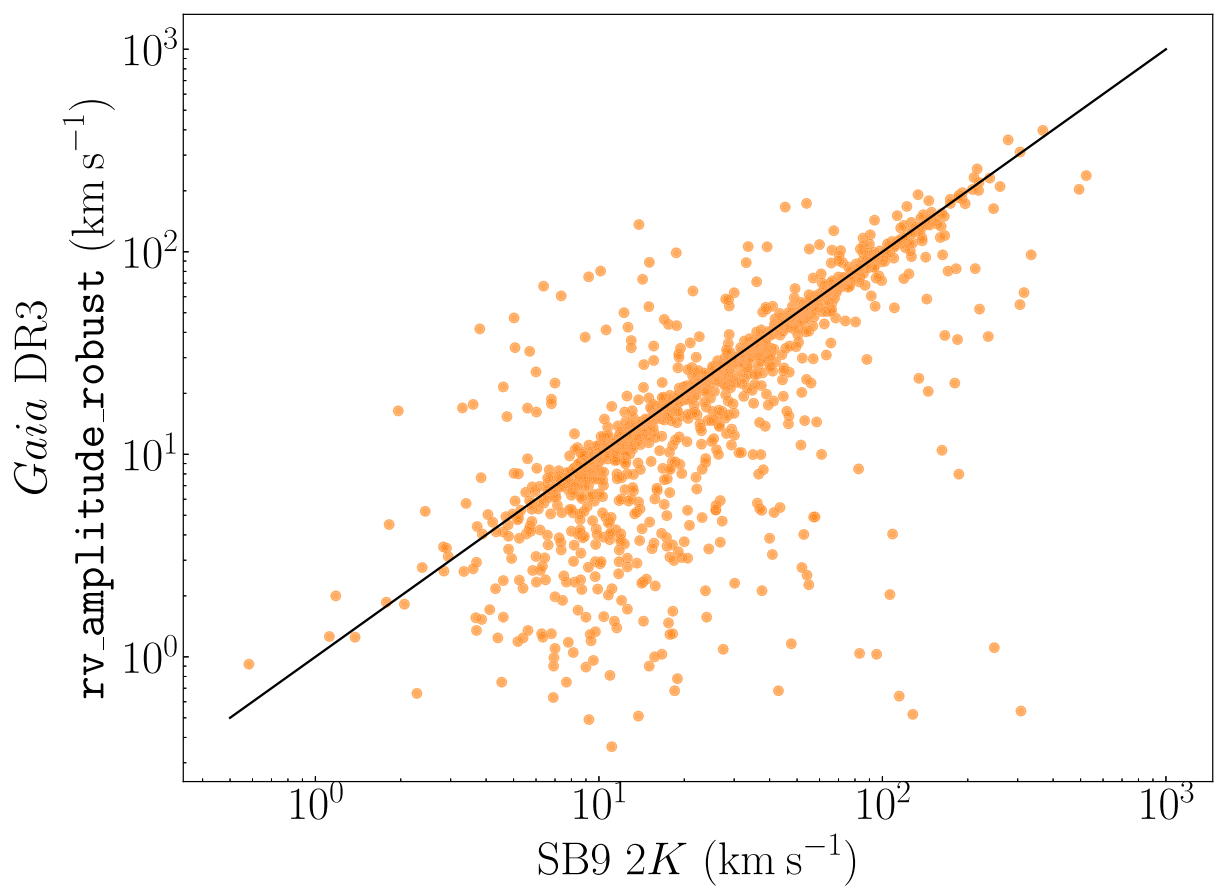

**Fig. 11.** Comparison between the `rv_amplitude_robust` values in the Gaia DR3 catalog and the RV full amplitudes $2K$ from the SB9 catalog (Pourbaix et al. 2004). The solid line corresponds to $2K =$ `rv_amplitude_robust`.

the actual radial velocity amplitude. This comparison shows that `rv_amplitude_robust` is generally a reliable proxy for $K$ for binary systems, but it does not necessarily imply that `rv_amplitude_robust` alone is effective for detecting binaries, i.e., the false positive rate is unknown.

## Appendix 2 RV data for observed targets

Tables 3 and 4 are the RV observations for J0144 and J2013, respectively.

**Table 3.** RV observations for J0144.

| MJD | RV (km s$^{-1}$) | Uncertainty (km s$^{-1}$) |
|---|---|---|
| 60528.77488 | 73.48 | 1.48 |
| 60529.67327 | 31.87 | 0.47 |
| 60529.80826 | 96.53 | 0.39 |
| 60530.81254 | −97.35 | 0.52 |
| 60537.81647 | −11.53 | 0.50 |
| 60579.52595 | 57.98 | 0.41 |
| 60579.60853 | 11.65 | 0.46 |
| 60580.49186 | 86.01 | 0.65 |
| 60581.54552 | −87.65 | 1.87 |
| 60582.49401 | −75.55 | 1.68 |
| 60583.51362 | 109.22 | 0.58 |
| 60583.63508 | 60.85 | 0.39 |

**Table 4.** Same as table 3, but for J2013.

| MJD | RV (km s$^{-1}$) | Uncertainty (km s$^{-1}$) |
|---|---|---|
| 60528.63208 | −120.64 | 0.18 |
| 60528.75403 | −110.97 | 0.39 |
| 60530.65659 | −43.29 | 0.13 |
| 60530.76688 | −66.24 | 0.18 |
| 60578.44390 | −104.04 | 0.14 |
| 60578.45905 | −102.67 | 0.07 |
| 60579.51561 | 62.14 | 0.08 |
| 60579.62249 | 60.70 | 0.21 |
| 60580.43337 | −73.00 | 0.25 |
| 60580.55358 | −93.77 | 0.24 |
| 60581.41038 | −70.98 | 0.08 |
| 60581.45711 | −62.48 | 0.29 |
| 60581.58530 | −36.62 | 0.44 |
| 60582.40179 | 58.04 | 0.83 |
| 60582.50831 | 50.92 | 0.08 |
| 60583.50669 | −117.06 | 0.21 |

## Appendix 3 Posterior distributions of our analyses

Figures 12 and 13 show the posterior distributions of the atmospheric parameters. Figures 14 and 15 show those for the isochrone fitting. Figures 16 and 17 show those for the RV fitting.

## Appendix 4 Photometric variation

In this section, we briefly discuss the compatibility of TESS light curves with ellipsoidal modulation and/or rotational modulation due to spots, given the orbital parameters. Figures 18 and 19 show, for each of the two targets, the TESS light curve and their Lomb–Scargle periodogram in each TESS sector.

Tables 5 and 6 show the variation of photometric periods $P_{\rm LC}$ for J0144 and J2013 among different sectors. The uncertainty of $P_{\rm LC}$ in each sector is obtained by performing $10^4$ bootstrap iterations to calculate the Lomb–Scargle periodogram of each sector. For both objects, $P_{\rm LC}$ varies slightly from sector to sector. The difference between $P_{\rm orb}$ and $(2\times)P_{\rm LC}$ is comparable to or larger than the statistical uncertainties, but less than the variation of $P_{\rm LC}$ among sectors. Thus $P_{\rm LC}$ is consistent with the assumption of synchronization.

Figure 3 shows the synthetic light curves calculated with `PHOEBE` (Prša & Zwitter 2005; Prsa et al. 2011; Prša 2018) using the binary parameters obtained by our analysis. For J2013, the amplitude of light modulation is roughly consistent. However, the phase of the light curve is inconsistent. Also, the heights of the two local maxima in the TESS light curve are different to an extent that the `PHOEBE` model does not expect. We interpreted these factors as the effect of stellar spots. The light curve of J0144 looks completely inconsistent with the `PHOEBE` ellipsoidal model, so we consider that the modulation by stellar spots dominates the ellipsoidal modulation in this system.

Following equations (3) and (4) in Notsu et al. (2019), we estimated the temperature $T_s$ of the spots and the surface area $A_s$ covered by them. Then the areal covering fraction of the spots $A_s/A_{\rm MS}$ is obtained using the relative drop in flux due to the spots $\delta F/F$. Here, $\delta F/F$ is calculated as the difference between the minimum and maximum of the residuals of the TESS light curve from the `PHOEBE` ellipsoidal model.

The resulting parameters of the spots are listed in tables 7 and 8. The spots cover 6%–13% (J0144) and 1%–2% (J2013) of the visible surface of the star, equivalent to 1.0–$2.2 \times 10^{17}$ m$^2$ for J0144 and 3.3–$4.7 \times 10^{16}$ m$^2$ for J2013. These values are among the typical spot coverage for solar-type stars with rotational periods between 0.24 and 11.16 d, which spans from $1 \times 10^{15}$ to $1 \times 10^{18}$ m$^2$ (Doyle et al. 2020). Similar sizes of spots are also observed in previously known MS-compact object systems such as CPD-65 264 (Hernandez et al. 2022b).

In conclusion, J0144 shows rotational modulation due to stellar spots which smears out the ellipsoidal modulation, and J2013 shows spotted ellipsoidal modulation, as observed in previously known systems. The spots are the typical ones for solar-type stars.

## Appendix 5 Photometric data

Tables 9 and 10 show the photometric data used in figure 6.

## Appendix 6 Gaia DR3 NSS catalog orbital solution

One of our targets, J2013, is reported as a single-lined spectroscopic binary (SB1) in Gaia NSS catalog. Table 11 shows the NSS orbital solution for J2013.

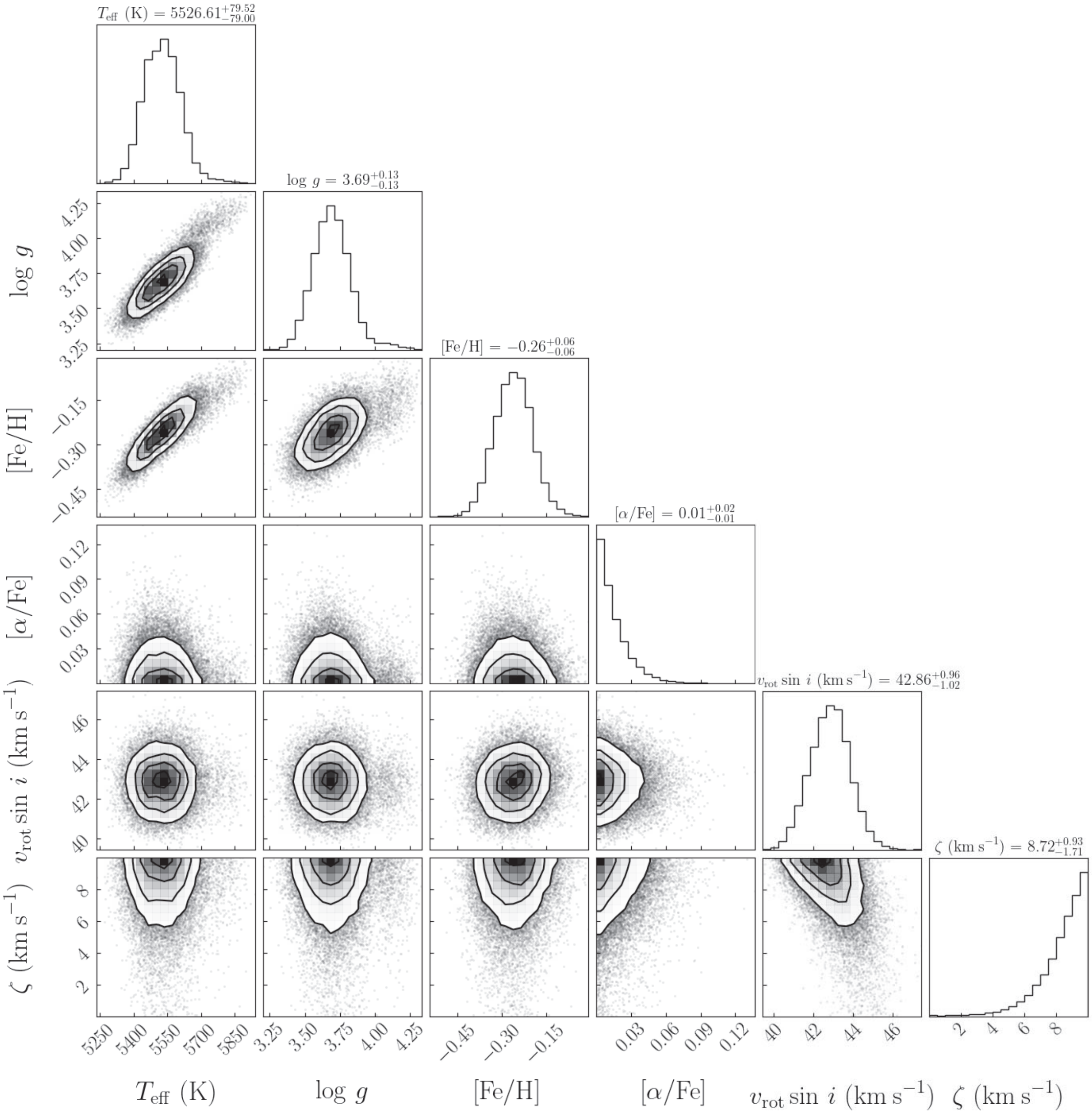

**Fig. 12.** Posterior distribution of the atmospheric parameters for J0144 derived by spectrum fitting (section 3).

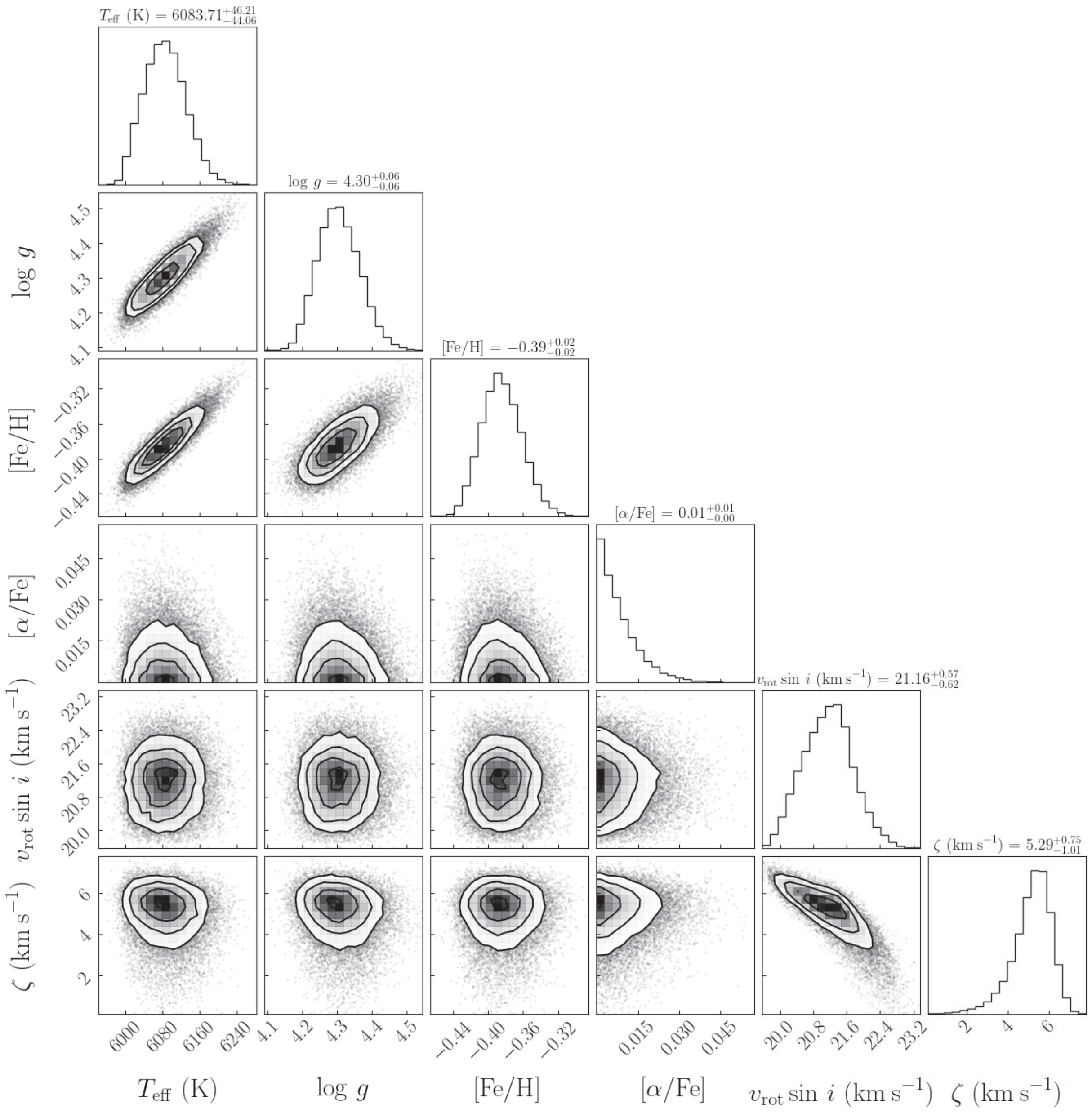

**Fig. 13.** Same as figure 12, but for J2013.

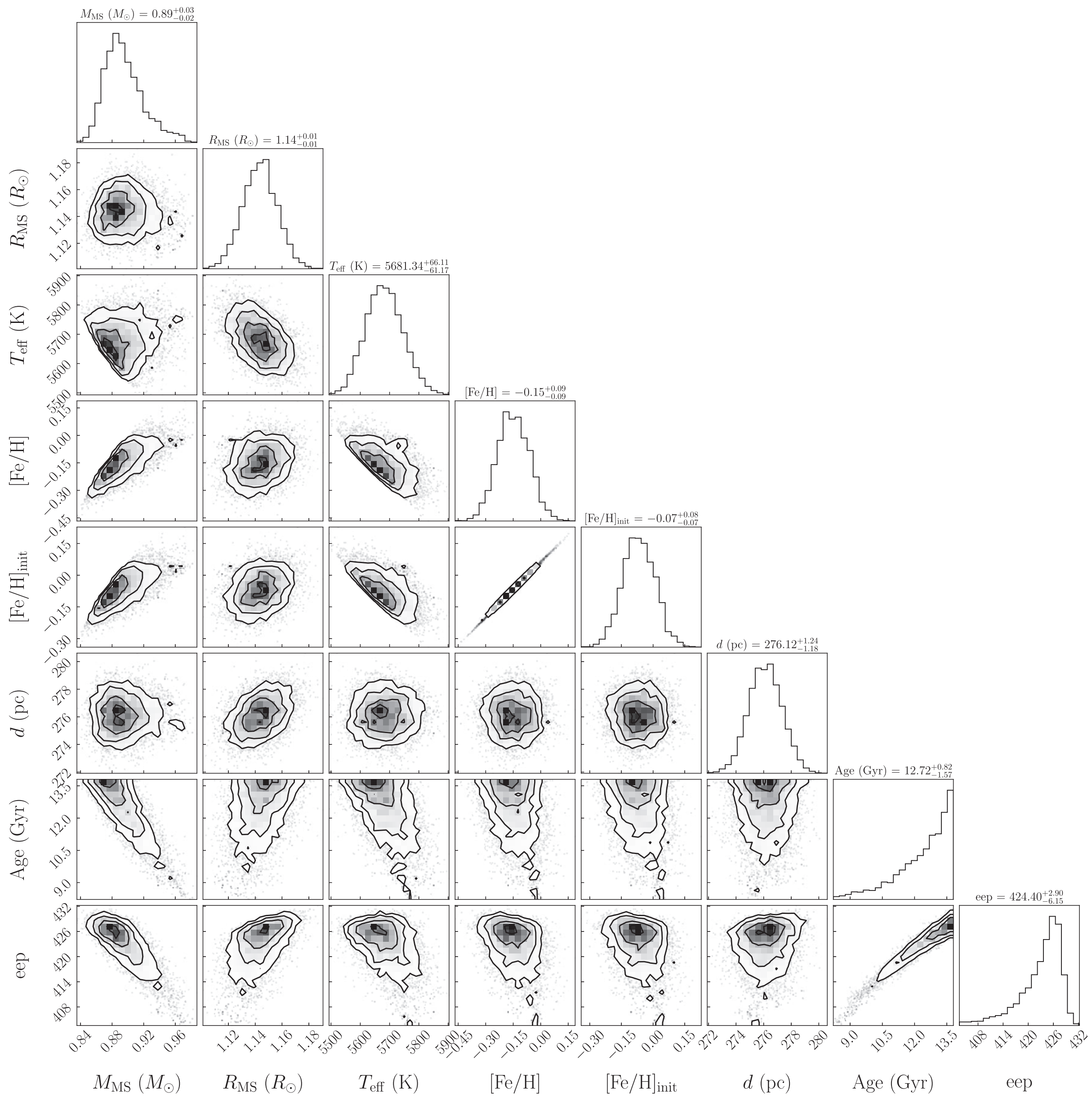

**Fig. 14.** Posterior distribution of parameters of the primary star for J0144 derived by isochrone fitting (subsection 4.1).

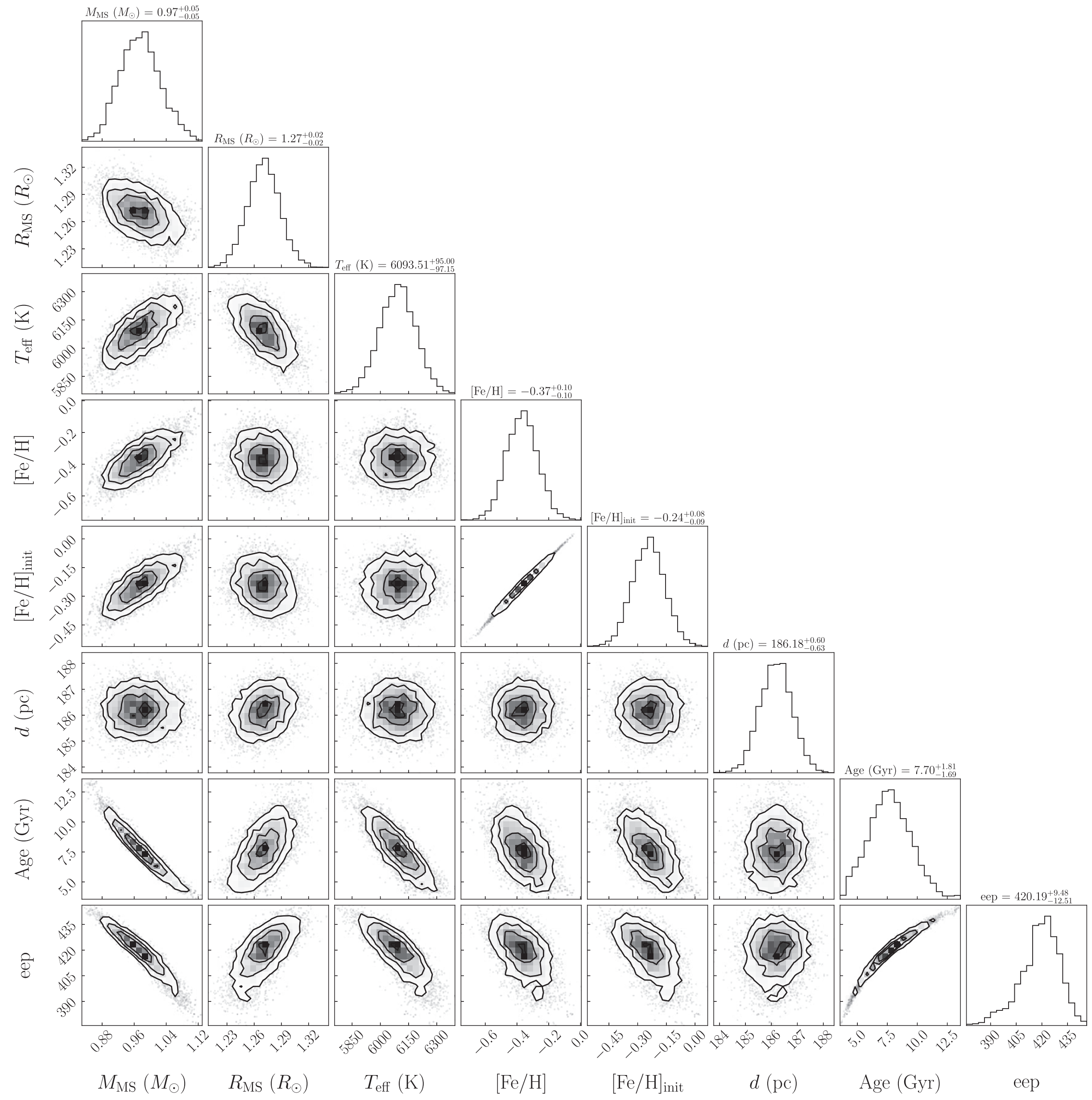

**Fig. 15.** Same as figure 14, but for J2013.

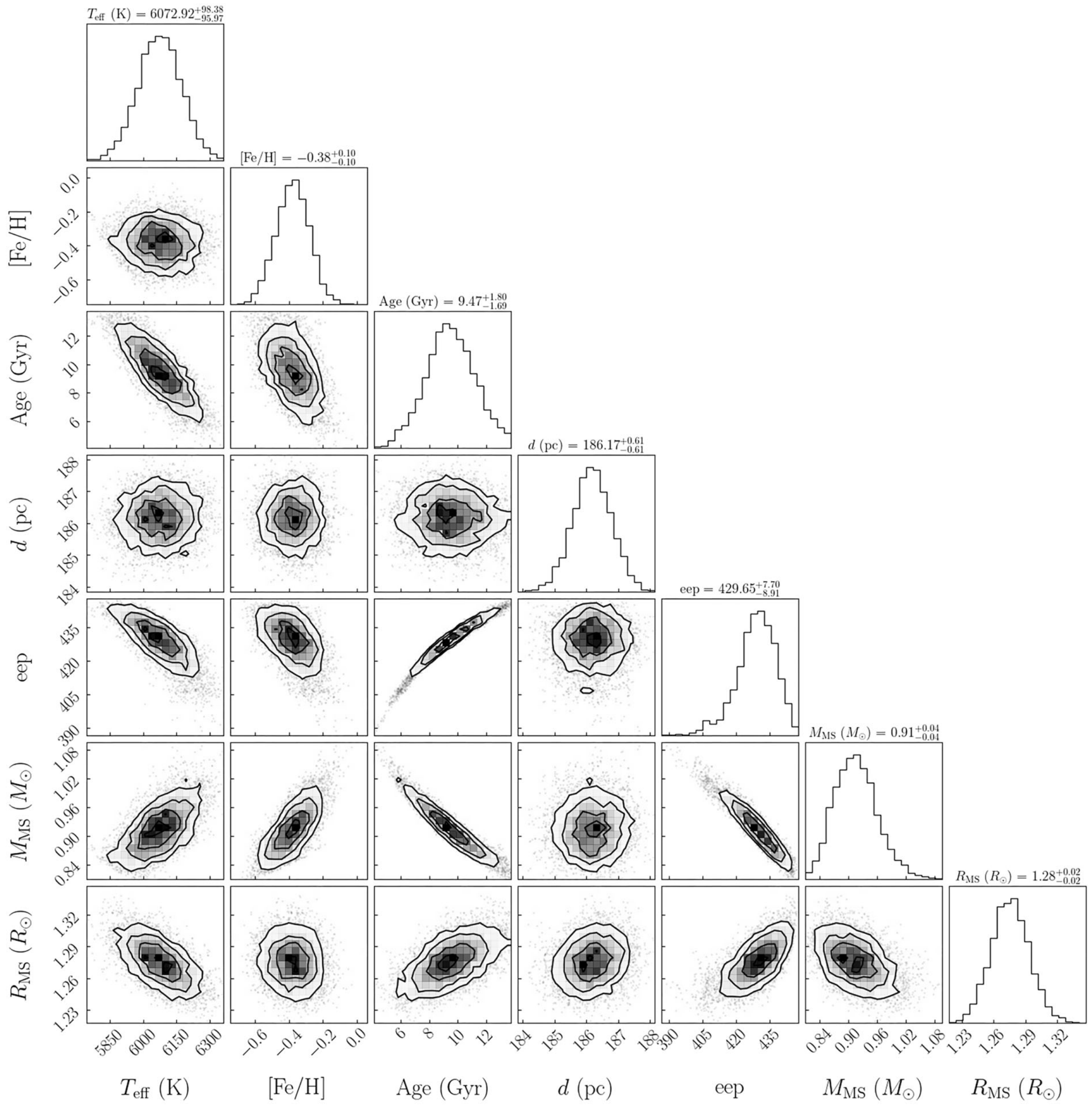

**Fig. 16.** Posterior distribution of orbital parameters for J0144 derived by RV fitting (subsection 4.2).

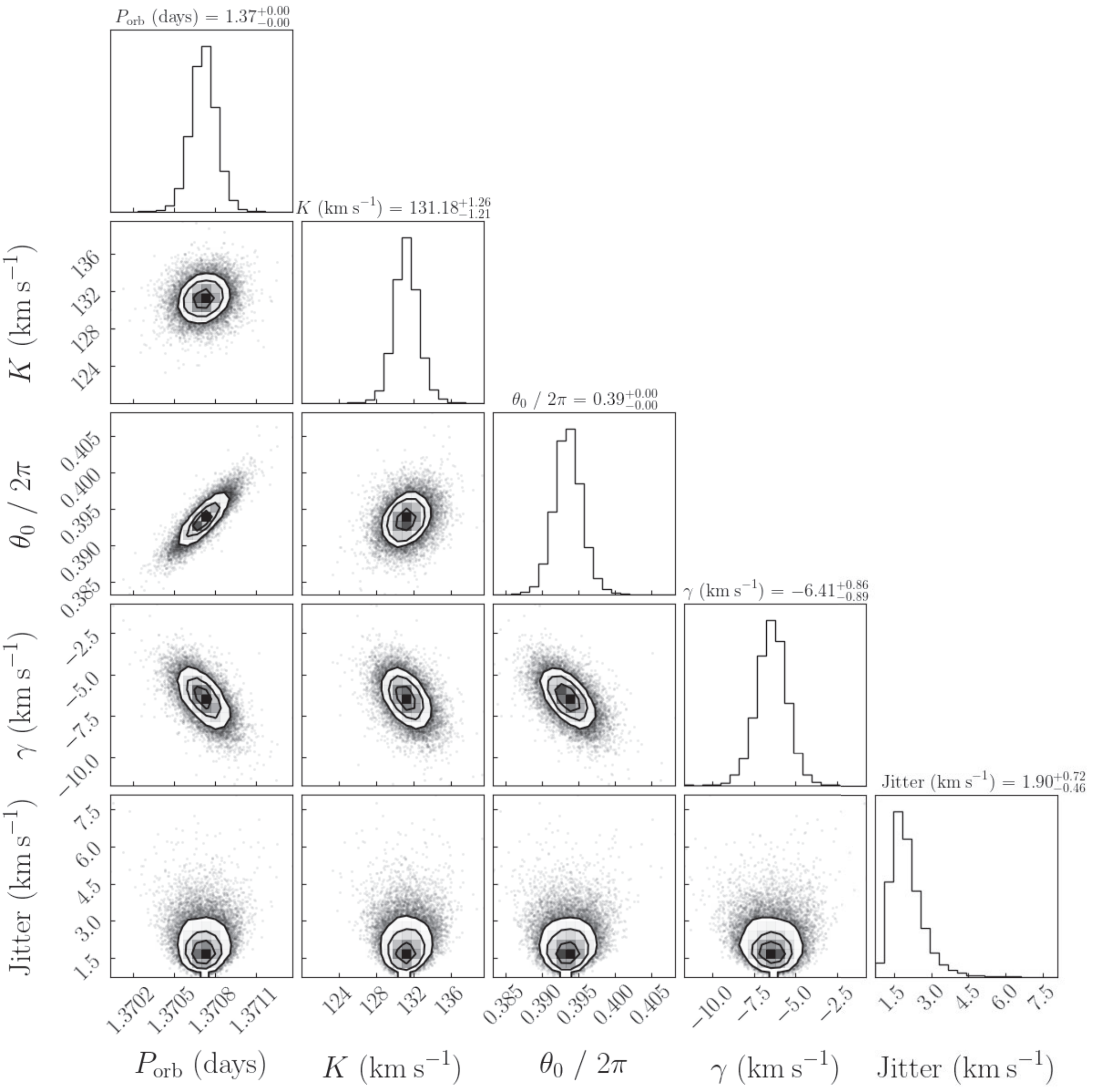

**Fig. 17.** Same as figure 16, but for J2013.

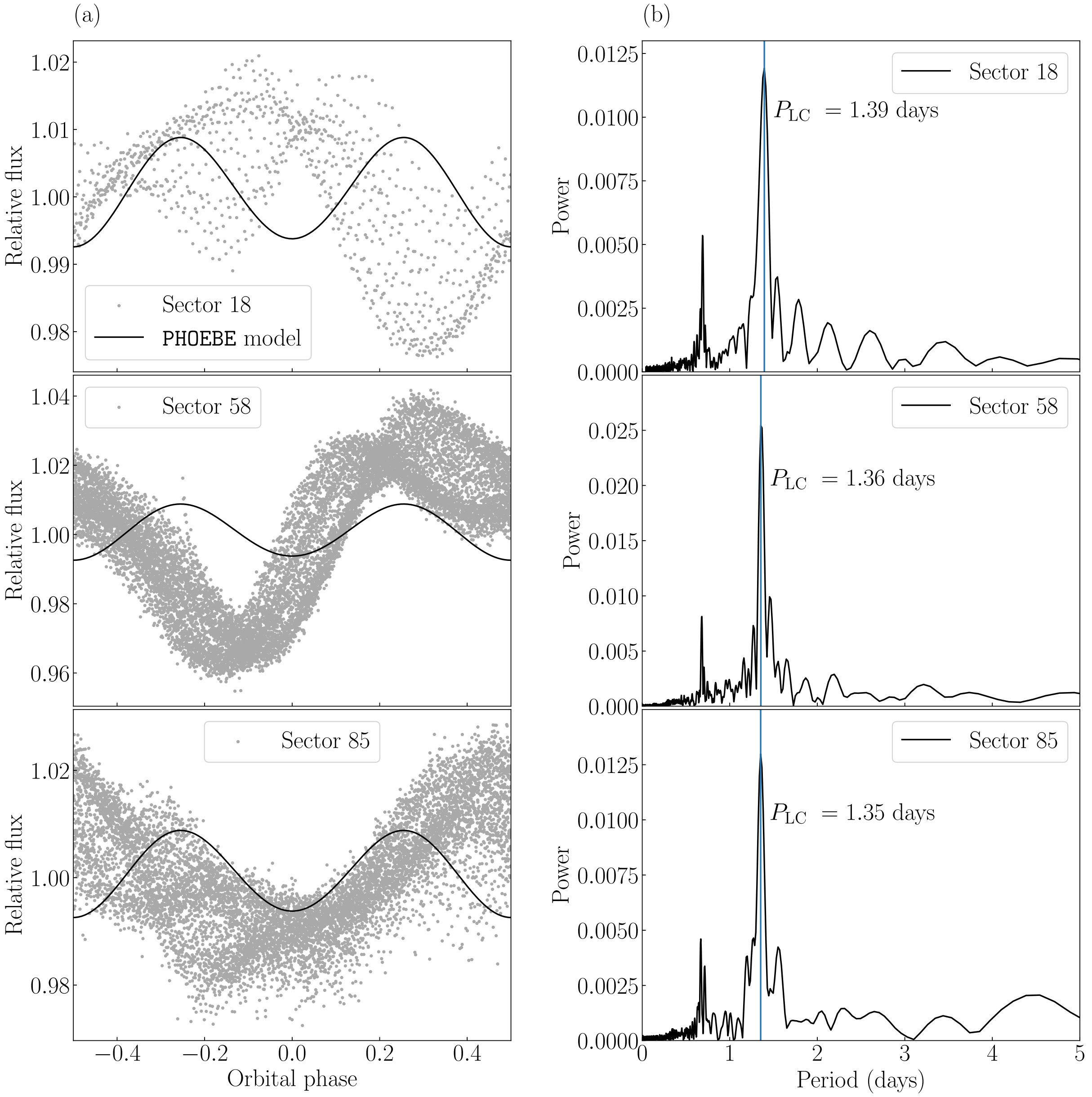

**Fig. 18.** (a) TESS light curve of J0144 in each sector. Each panel shows the folded data with the orbital period $P_{\rm orb}$ obtained by RV fitting. Also depicted are the PHOEBE model light curves with ellipsoidal modulation using the stellar and orbital parameters of J0144 obtained in our analysis. (b) Lomb–Scargle periodogram for TESS light curve of J0144 in each sector.

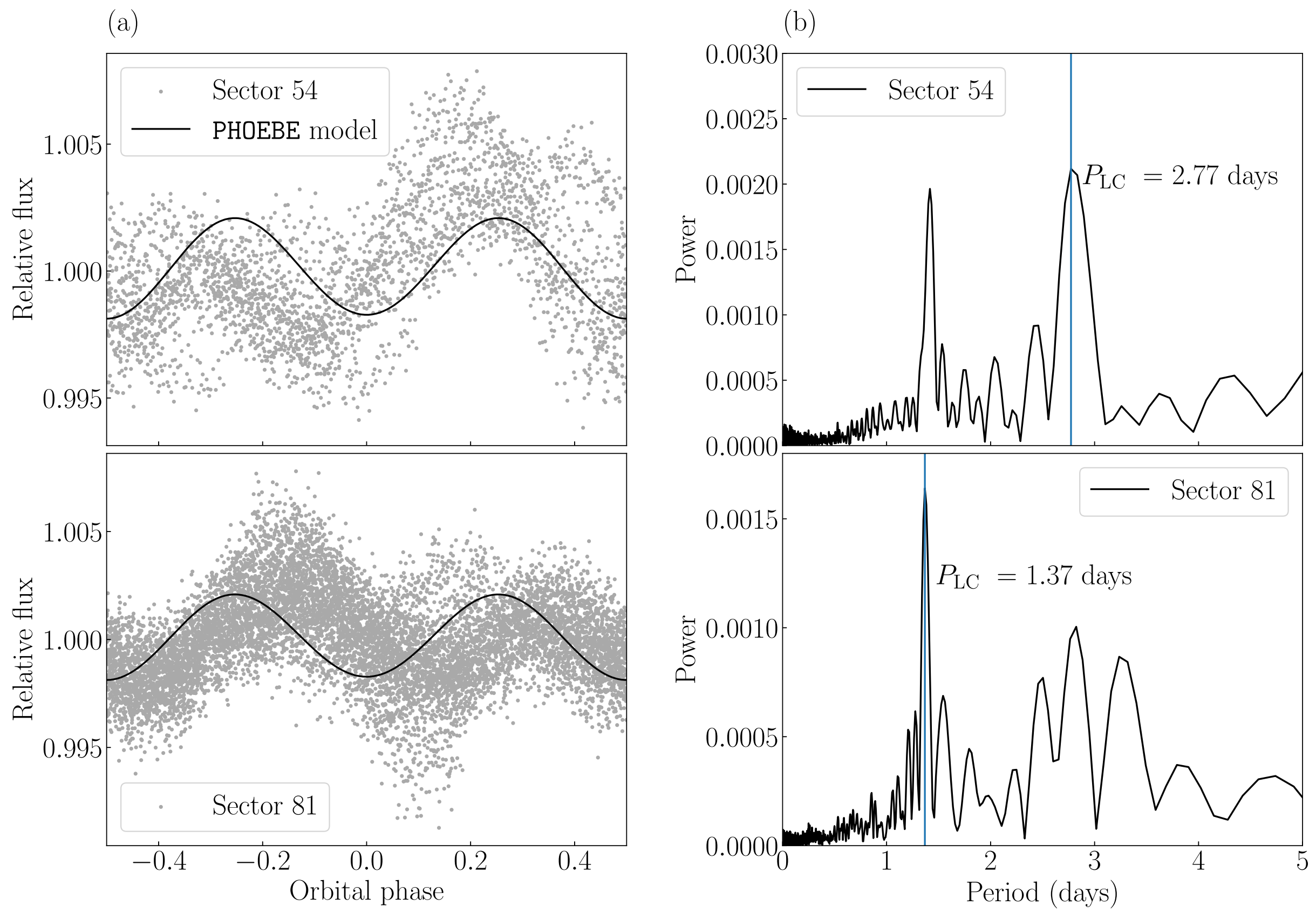

**Fig. 19.** Same as figure 18, but for J2013.

**Table 5.** Period of photometric variation for J0144.

| | | Sector | | |
|---|---|---|---|---|
| | | 18 | 58 | 85 |
| Orbital period | $P_{orb}$ (d) | | $1.370714 \pm 0.000090$ | |
| Photometric period | $P_{LC}$ (d) | $1.393 \pm 0.002$ | $1.355 \pm 0.002$ | $1.3529 \pm 0.0003$ |

**Table 6.** Same as table 5, but for J2013.

| | | Sector | |
|---|---|---|---|
| | | 54 | 81 |
| Orbital period | $P_{orb}$ (d) | $2.75736 \pm 0.00012$ | |
| Photometric period | $P_{LC}$ (d) | $2.77 \pm 0.01$ | $1.36800 \pm 0.00016$ |

**Table 7.** Parameters of spots for J0144.

| | | Sector | | |
|---|---|---|---|---|
| | | 18 | 58 | 85 |
| Spot temperature | $T_s$ (K) | | 3900 | |
| Flux drop | $\delta F/F$ | 4.8% | 10.0% | 7.0% |
| Spot coverage | $A_s/A_{MS}$ | 6.0% | 12.4% | 8.8% |

**Table 8.** Same as table 7, but for J2013.

| | | Sector | |
|---|---|---|---|
| | | 54 | 81 |
| Spot temperature | $T_s$ (K) | 4000 | |
| Flux drop | $\delta F/F$ | 1.3% | 1.9% |
| Spot coverage | $A_s/A_{MS}$ | 1.7% | 2.4% |

**Table 9.** Photometric observations of J0144.*

| Passband | Magnitude | System | Reference |
|---|---|---|---|
| Gaia DR3 *BP* | $12.079 \pm 0.004$ | Vaga | (1) |
| Gaia DR3 *G* | $11.725 \pm 0.003$ | Vega | (1) |
| Gaia DR3 *RP* | $11.203 \pm 0.004$ | Vega | (1) |
| Pan-Starrs DR1 *g* | 12.189 | AB | (2) |
| Pan-Starrs DR1 *r* | 11.798 | AB | (2) |
| Pan-Starrs DR1 *i* | 11.625 | AB | (2) |
| Pan-Starrs DR1 *z* | 11.529 | AB | (2) |
| Pan-Starrs DR1 *y* | 11.493 | AB | (2) |
| 2MASS *J* | $10.580 \pm 0.023$ | Vega | (3) |
| 2MASS *H* | $10.333 \pm 0.033$ | Vega | (3) |
| 2MASS *K* | $10.234 \pm 0.019$ | Vega | (3) |
| WISE *W1* | $10.171 \pm 0.023$ | Vega | (4) |
| WISE *W2* | $10.184 \pm 0.020$ | Vega | (4) |
| WISE *W3* | $9.908 \pm 0.055$ | Vega | (4) |

*References: (1) Gaia Collaboration (2023a); (2) Chambers et al. (2016); (3) Skrutskie et al. (2006); and (4) Cutri et al. (2021).

**Table 10.** Same as table 9, but for J2013.*

| Passband | Magnitude | System | Reference |
|---|---|---|---|
| Gaia DR3 *BP* | $10.547 \pm 0.003$ | Vaga | (1) |
| Gaia DR3 *G* | $10.301 \pm 0.003$ | Vega | (1) |
| Gaia DR3 *RP* | $9.893 \pm 0.004$ | Vega | (1) |
| 2MASS *J* | $9.394 \pm 0.024$ | Vega | (2) |
| 2MASS *H* | $9.124 \pm 0.027$ | Vega | (2) |
| 2MASS *K* | $9.066 \pm 0.023$ | Vega | (2) |
| WISE *W1* | $8.905 \pm 0.023$ | Vega | (3) |
| WISE *W2* | $8.928 \pm 0.020$ | Vega | (3) |
| WISE *W3* | $8.925 \pm 0.027$ | Vega | (3) |

*References: (1) Gaia Collaboration (2023a); (2) Skrutskie et al. (2006); and (3) Cutri et al. (2021).

**Table 11.** The orbital solution for J2013 in the Gaia DR3 NSS catalog.

| | | J2013+1734 |
|---|---|---|
| Source ID | | 1809974711191044864 |
| Orbital period | $P_{\rm orb}$ (d) | $2.757598 \pm 2.8 \cdot 10^{-5}$ |
| Peri astron passage | $t_{\rm peri} -$ J2016.0 (d) | $0.99 \pm 0.18$ |
| Eccentricity | $e$ | $0.0108. \pm 0.0063$ |
| Center of mass velocity | $\gamma$ (km s$^{-1}$) | $-30.93 \pm 0.33$ |
| RV semi-amplitude | $K$ (km s$^{-1}$) | $92.86 \pm 0.51$ |
| Argument of periastron | (°) | $60.7 \pm 23.4$ |
| Mass function | ($M_\odot$) | 0.224 |
| n_obs | | 20 |
| Goodness of fit | | 0.44 |
| Significance | | 183.6 |

## References

Abdul-Masih, M., et al. 2020, Nature, 580, E11
Aigrain, S., Pont, F., & Zucker, S. 2012, MNRAS, 419, 3147
Astropy Collaboration 2013, A&A, 558, A33
Astropy Collaboration 2018, AJ, 156, 123
Astropy Collaboration 2022, ApJ, 935, 167
Betancourt, M. 2017, arXiv:1701.02434
Bianchi, L., Hutchings, J., Bohlin, R., Thilker, D., & Berti, E. 2024, ApJ, 976, 131
Bingham, E., et al. 2018, arXiv:1810.09538
Bodensteiner, J., et al. 2020, A&A, 641, A43
Breton, R. P., Rappaport, S. A., van Kerkwijk, M. H., & Carter, J. A. 2012, ApJ, 748, 115
Brown, W. R., Gianninas, A., Kilic, M., Kenyon, S. J., & Allende Prieto, C. 2016, ApJ, 818, 155
Buchner, J. 2016, Astrophys. Source Code Library, ascl:1606.005
Buchner, J., et al. 2014, A&A, 564, A125
Carter, J. A., Rappaport, S., & Fabrycky, D. 2011, ApJ, 728, 139
Chakrabarti, S., et al. 2023, AJ, 166, 6
Chambers, K. C., et al. 2016, arXiv:1612.05560
Choi, J., Dotter, A., Conroy, C., Cantiello, M., Paxton, B., & Johnson, B. D. 2016, ApJ, 823, 102
Clayton, G. C. 2012, J. Am. Asso. Variable Star Obs., 40, 539
Coelho, P., Barbuy, B., Meléndez, J., Schiavon, R. P., & Castilho, B. V. 2005, A&A, 443, 735
Cutri, R. M., et al. 2021, VizieR Online Data Catalog, II/328
Davis, P. J., Kolb, U., & Willems, B. 2010, MNRAS, 403, 179
Ding, X., et al. 2024, AJ, 168, 217
Dittmann, A. 2024, Open J. Astrophys., 7, 79
Dotter, A. 2016, ApJS, 222, 8
Doyle, L., Ramsay, G., & Doyle, J. G. 2020, MNRAS, 494, 3596
Duane, S., Kennedy, A. D., Pendleton, B. J., & Roweth, D. 1987, Phys. Lett. B, 195, 216
El-Badry, K., et al. 2023a, MNRAS, 518, 1057
El-Badry, K., et al. 2023b, MNRAS, 521, 4323
El-Badry, K., et al. 2024a, Open J.Astrophys., 7, 27
El-Badry, K., et al. 2024b, Open J.Astrophys., 7, 58
El-Badry, K., & Quataert, E. 2021, MNRAS, 502, 3436
El-Badry, K., & Rix, H.-W. 2022, MNRAS, 515, 1266
El-Badry, K., Seeburger, R., Jayasinghe, T., Rix, H.-W., Almada, S., Conroy, C., Price-Whelan, A. M., & Burdge, K. 2022, MNRAS, 512, 5620
Faigler, S., Kull, I., Mazeh, T., Kiefer, F., Latham, D. W., & Bloemen, S. 2015, ApJ, 815, 26
Feroz, F., & Hobson, M. P. 2008, MNRAS, 384, 449
Feroz, F., Hobson, M. P., & Bridges, M. 2009, MNRAS, 398, 1601
Feroz, F., Hobson, M. P., Cameron, E., & Pettitt, A. N. 2019, Open J. Astrophys., 2, 10
Ferrario, L. 2025, arXiv:2503.00411
Fitzpatrick, M., Placco, V., Bolton, A., Merino, B., Ridgway, S., & Stanghellini, L. 2025, in ASP Conf. Ser., 541, Astronomical Data Analysis Software and Systems XXXIII, ed. A. Jacques et al. (San Francisco: ASP), 461
Fulton, B. J., & Petigura, E. A. 2018, AJ, 156, 264
Gaia Collaboration 2023a, A&A, 674, A1
Gaia Collaboration 2023b, A&A, 674, A34
Gaia Collaboration 2024, A&A, 686, L2
Gelman, A., Carlin, J. B., Stern, H. S., Dunson, D. B., Vehtari, A., & Rubin, D. B. 2013, Bayesian Data Analysis, 3rd ed.(New York: Chapman and Hall/CRC)
Green, G. M., Schlafly, E., Zucker, C., Speagle, J. S., & Finkbeiner, D. 2019, ApJ, 887, 93
Green, M. J., et al. 2025, A&A, 695, A210
Hernandez, M. S., et al. 2021, MNRAS, 501, 1677
Hernandez, M. S., et al. 2022a, MNRAS, 512, 1843
Hernandez, M. S., et al. 2022b, MNRAS, 517, 2867
Hoffman, M., Blei, D. M., Wang, C., & Paisley, J. 2012, arXiv:1206.7051
Huang, C. X., et al. 2020a, Res. Notes AAS, 4, 204
Huang, C. X., et al. 2020b, Res. Notes AAS, 4, 206
Iben, I., Tutukov, A. V., & Yungelson, L. R. 1996, ApJ, 456, 750
Irrgang, A., Geier, S., Kreuzer, S., Pelisoli, I., & Heber, U. 2020, A&A, 633, L5
Jayasinghe, T., et al. 2021, MNRAS, 504, 2577
Jayasinghe, T., et al. 2022, MNRAS, 516, 5945
Kato, M., & Hachisu, I. 1989, ApJ, 346, 424
Kawahara, H., Masuda, K., MacLeod, M., Latham, D. W., Bieryla, A., & Benomar, O. 2018, AJ, 155, 144
Kilic, M., et al. 2014, MNRAS, 438, L26
Kiziltan, B., Kottas, A., De Yoreo, M., & Thorsett, S. E. 2013, ApJ, 778, 66
Kruse, E., & Agol, E. 2014, Science, 344, 275
Krushinsky, V., et al. 2020, MNRAS, 493, 5208
Kunimoto, M., et al. 2021, Res. Notes AAS, 5, 234
Kunimoto, M., Tey, E., Fong, W., Hesse, K., Shporer, A., Fausnaugh, M.,

Vanderspek, R., & Ricker, G. 2022, Res. Notes AAS, 6, 236
Landsman, W., Simon, T., & Bergeron, P. 1993, PASP, 105, 841
Lejeune, T., Cuisinier, F., & Buser, R. 1997, A&AS, 125, 229
Lejeune, T., Cuisinier, F., & Buser, R. 1998, A&AS, 130, 65
Li, X., Wang, S., Zhao, X., Bai, Z., Yuan, H., Zhang, H., & Liu, J. 2022, ApJ, 938, 78
Lin, J., et al. 2023, ApJ, 944, L4
Lin, J., Rappaport, S., Podsiadlowski, P., Nelson, L., Paxton, B., & Todorov, P. 2011, ApJ, 732, 70
Liu, H.-B., Gu, W.-M., Zhang, Z.-X., Yi, T., Liu, J.-Z., & Sun, M. 2024, ApJ, 969, 114
Lomb, N. R. 1976, Ap&SS, 39, 447
Marsh, T. R., Nelemans, G., & Steeghs, D. 2004, MNRAS, 350, 113
Masuda, K. 2022, ApJ, 937, 94
Masuda, K., & Hotokezaka, K. 2019, ApJ, 883, 169
Masuda, K., Kawahara, H., Latham, D. W., Bieryla, A., Kunitomo, M., MacLeod, M., & Aoki, W. 2019, ApJ, 881, L3
Maxted, P. F. L., et al. 2013, Nature, 498, 463
Mazeh, T., et al. 2022, MNRAS, 517, 4005
Muirhead, P. S., et al. 2013, ApJ, 767, 111
Nomoto, K., & Kondo, Y. 1991, ApJ, 367, L19
Nomoto, K., & Leung, S.-C. 2018, Space Sci. Rev., 214, 67
Nomoto, K., Thielemann, F.-K., & Yokoi, K. 1984, ApJ, 286, 644
Notsu, Y., et al. 2019, ApJ, 876, 58
O'Brien, M. S., Bond, H. E., & Sion, E. M. 2001, ApJ, 563, 971
Ochsenbein, F., Bauer, P., & Marcout, J. 2000, A&AS, 143, 23
Paczyński, B. 1967, Acta Astron., 17, 287
Paegert, M., Stassun, K. G., Collins, K. A., Pepper, J., Torres, G., Jenkins, J., Twicken, J. D., & Latham, D. W. 2021, arXiv:2108.04778
Parsons, S. G., et al. 2015, MNRAS, 452, 1754
Parsons, S. G., et al. 2023, MNRAS, 518, 4579
Parsons, S. G., Rebassa-Mansergas, A., Schreiber, M. R., Gänsicke, B. T., Zorotovic, M., & Ren, J. J. 2016, MNRAS, 463, 2125
Paxton, B., et al. 2013, ApJS, 208, 4
Paxton, B., et al. 2015, ApJS, 220, 15
Paxton, B., Bildsten, L., Dotter, A., Herwig, F., Lesaffre, P., & Timmes, F. 2011, ApJS, 192, 3
Perets, H. B., et al. 2010, Nature, 465, 322
Phan, D., Pradhan, N., & Jankowiak, M. 2019, arXiv:1912.11554
Pietrzyński, G., et al. 2012, Nature, 484, 75
Pourbaix, D., et al. 2004, A&A, 424, 727
Prsa, A., Matijevic, G., Latkovic, O., Vilardell, F., & Wils, P. 2011, Astrophys. Source Code Library, ascl:1106.002
Prša, A. 2018, Modeling and Analysis of Eclipsing Binary Stars (Bristol: IOP Publishing)
Prša, A., & Zwitter, T. 2005, ApJ, 628, 426
Qi, S., Gu, W.-M., Yi, T., Zhang, Z.-X., Wang, S., & Liu, J. 2023, AJ, 165, 187
Rappaport, S., Nelson, L., Levine, A., Sanchis-Ojeda, R., Gandolfi, D., Nowak, G., Palle, E., & Prsa, A. 2015, ApJ, 803, 82
Rebassa-Mansergas, A., et al. 2017, MNRAS, 472, 4193
Rebassa-Mansergas, A., Gänsicke, B. T., Schreiber, M. R., Koester, D., & Rodríguez-Gil, P. 2010, MNRAS, 402, 620
Rebassa-Mansergas, A., Ren, J. J., Parsons, S. G., Gänsicke, B. T., Schreiber, M. R., García-Berro, E., Liu, X.-W., & Koester, D. 2016, MNRAS, 458, 3808
Rowan, D. M., et al. 2024, MNRAS, 529, 587
Sato, B., et al. 2024, in Proc. SPIE, 13096, Ground-based and Airborne Instrumentation for Astronomy X, ed. J. J. Bryant et al. (Bellingham, WA: SPIE), 1309644
Scargle, J. D. 1982, ApJ, 263, 835
Schlafly, E. F., et al. 2016, ApJ, 821, 78
Shahaf, S., Hallakoun, N., Mazeh, T., Ben-Ami, S., Rekhi, P., El-Badry, K., & Toonen, S. 2024, MNRAS, 529, 3729
Shen, K. J. 2015, ApJ, 805, L6
Skrutskie, M. F., et al. 2006, AJ, 131, 1163
Tanikawa, A., Hattori, K., Kawanaka, N., Kinugawa, T., Shikauchi, M., & Tsuna, D. 2023, ApJ, 946, 79
Thompson, T. A., et al. 2019, Science, 366, 637
Tody, D. 1986, in Proc. SPIE, 627, Instrumentation in Astronomy VI, ed. D. L. Crawford, (Bellingham:SPIE), 733
Tody, D. 1993, in ASP Conf. Ser., 52, Astronomical Data Analysis Software and Systems II, ed. R. J. Hanisch (San Francisco: ASP), 173
Tomoyoshi, M., et al. 2024, ApJ, 977, 151
Tucker, M. A., Wheeler, A. J., Rowan, D. M., & Huber, M. E. 2025, Open J. Astrophys., 8, 61
van Kerkwijk, M. H., Rappaport, S. A., Breton, R. P., Justham, S., Podsiadlowski, P., & Han, Z. 2010, ApJ, 715, 51
Wang, S., et al. 2024, Nature Astron., 8, 1583
Webbink, R. F. 1984, ApJ, 277, 355
Wonnacott, D., Kellett, B. J., & Stickland, D. J. 1993, MNRAS, 262, 277
Yamaguchi, N., et al. 2024a, MNRAS, 527, 11719
Yamaguchi, N., El-Badry, K., Ciardi, D. R., Latham, D. W., Masuda, K., Bieryla, A., Clark, C. A., & Condon, S. S. 2024b, PASP, 136, 074201
Yamaguchi, N., El-Badry, K., Rees, N. R., Shahaf, S., Mazeh, T., & Andrae, R. 2024c, PASP, 136, 084202
Yi, T., et al. 2022, Nature Astron., 6, 1203
Yuan, H., et al. 2022, ApJ, 940, 165
Zhang, X. B., Fu, J. N., Liu, N., Luo, C. Q., & Ren, A. B. 2017, ApJ, 850, 125
Zhang, Z.-X., Liu, H.-B., Yi, T., Sun, M., & Gu, W.-M. 2024, ApJ, 961, L48
Zhao, X., et al. 2024, ApJ, 964, 101
Zheng, L.-L., et al. 2022, ApJ, 936, 33
Zheng, L.-L., et al. 2023, Sci. China Phys. Mec. Astron., 66, 129512
Zhu, H., Wang, W., Li, X., Li, J., & Tian, P. 2025, J. High Energy Astrophys., 45, 428
Zorotovic, M., Schreiber, M. R., & Gänsicke, B. T. 2011, A&A, 536, A42
Zucker, S. 2003, MNRAS, 342, 1291